\documentclass{article}
\usepackage[utf8]{inputenc}
\usepackage{amsmath}
\usepackage{amsfonts}
\usepackage{amssymb}
\usepackage{caption}
\usepackage{float}
\usepackage{graphicx}
\usepackage{tikz}
\usepackage{mathtools}
\usepackage[margin=0.9in]{geometry}
\usepackage{calrsfs}
\usepackage{bbm}
\usepackage{enumitem}

\DeclareMathAlphabet{\pazocal}{OMS}{zplm}{F}{n}
\DeclareMathAlphabet{\pazocal}{OMS}{zplm}{N}{n}
\DeclareMathAlphabet{\pazocal}{OMS}{zplm}{B}{n}
\DeclareMathAlphabet{\pazocal}{OMS}{zplm}{D}{n}
\newcommand\subpsim{\mathrel{%
  \ooalign{\raise0.2ex\hbox{$p$}\cr\hidewidth\raise-0.8ex\hbox{\scalebox{0.9}{$\sim$}}\hidewidth\cr}}}
\newcommand\subnsim{\mathrel{%
  \ooalign{\raise0.2ex\hbox{$n$}\cr\hidewidth\raise-0.8ex\hbox{\scalebox{0.9}{$\sim$}}\hidewidth\cr}}}
\usepackage{indentfirst}
\usepackage{dirtytalk}
\usepackage[utf8]{inputenc}
\newcommand{\mat}[1]{\mbox{\boldmath{$#1$}}} 
\usepackage[caption=false]{subfig}
\usepackage{listings}
\usepackage{fancyhdr}

\newtheorem{definition}{Definition}[section]
\usepackage{enumitem}
\usepackage{enumitem}
\usepackage{diagbox}
\usepackage{booktabs}
\usepackage{subfig}
\allowdisplaybreaks
\usepackage{physics}
\providecommand{\keywords}[1]
{
  \small	
  \textbf{Keywords} #1
}
\usepackage{multirow}
\usepackage{authblk}

\usepackage{mathrsfs}
\usepackage{multirow}
\usepackage{physics}
\allowdisplaybreaks
\newtheorem{prop}{Proposition}
\newtheorem{lemma}{Lemma}
\usepackage{xcolor}

\title{Time-varying STARMA models by wavelets}
\author[1]{Yangyang Chen}
\author[1,*]{Pedro A. Morettin}

\author[1]{Chang Chiann.}

\affil[1]{Department of Statistics, University of São Paulo}
\date{}

\begin{document}

\maketitle

\let\thefootnote\relax\footnote{*Corresponding author.}
\let\thefootnote\relax\footnote{E-mail address: pam@ime.usp.br (P. A. Morettin).}

\begin{abstract}
    The spatio-temporal autoregressive moving average (STARMA) model is frequently used in several studies of multivariate time series data, where the assumption of stationarity is important, but it is not always guaranteed in practice. One way to proceed is to consider locally stationary processes. In this paper we propose a time-varying spatio-temporal autoregressive and moving average (tvSTARMA) modelling based on the locally stationarity assumption. The time-varying parameters are expanded as linear combinations of wavelet bases and procedures are proposed to estimate the coefficients. Some simulations and an application to historical daily precipitation records of Midwestern states of the USA are illustrated.
\end{abstract}


\section{Introduction}

Since Box and Jenkins (1970) introduced the class of seasonal autoregressive integrated moving average (SARIMA) model as a procedure for identifying, estimating, and checking models for a specific time series data set, the method became standard for analyzing stationary and homogeneous nonstationary time series with constant coefficients and variance. The models are fitted to time series data either to better understand the data or to predict future points in the series (forecasting). In the case of multivariate time series that are stationary, models of the vector autoregressive and moving average (VARMA) family are often used.

In the space-time geostatistical context, the classic model is the STARMA (Spatio-Temporal Autoregressive and Moving Average) model which is a special case of the VARMA model.
Cliff and Ord (1975) and Martin and Oeppen (1975) were the first to use models of the class STARMA and then, several methods were developed by Pfeifer and Deutsch (1980a, 1980b, 1981a, 1981b, 1981c) at the early eighties, but it went through a period of oblivion due to the difficulty of computational implementation. Until recent years, the STARMA model has been revived and is often used in geostatistical analyzes to describe spatially localized time series characterized by linear dependence lagged in both space and time, that is, the modeling processes are characterized by random variables observed at $N$ geographic locations and, at each location, $T$ observations over time.

In this case, in addition to the fact that recent past values have more influence, close locations also have more influence than distant locations, through the specification of spatial weight matrices which give the highest weights to the nearest neighbors.
The STARMA model has already been widely applied for different types of space time data for example, real estate price (Pace et al. (2000)), traffic flow data (Kamarianakis and Prastacos (2005)), damage detection (Hu et al. (2011)), regional bank deposits (Kurt and Tunay (2015)), wind power (Zou et al (2018)), etc.

As the models of class ARIMA, the STARMA models also have the assumption of stationarity, however, in practice, this assumption is not warranted.
There are many well known techniques that convert nonstationary time series into stationary ones, such as differencing, log transformation, detrending, etc. An attractive alternative is the idea of a locally stationary process in which the process is approximately stationary over small periods of time, but whose characteristics (covariances, parameters, etc.) are gradually changing throughout the time period. Many estimators and asymptotic results were developed by Dahlhaus (1996a, 1996b, 1996c, 1997, 2000) and Dahlhaus et al. (1999) and an overview of the locally stationary process can be found in Dalhaus (2012). 

Time-varying models based on the locally stationary process introduced by Dahlhaus have been used over the years in many different cases, for example, Chiann and Morettin (1999, 2005) investingated the estimation of time varying coefficients of a linear system, Sato et al. (2007) proposed an estimation procedure for time-varying VAR models and applied to functional magnetic resonance imaging dates, Rohan and Ramanathan (2013) introduced a nonparametric estimation to time-varying Generalized Autoregressive Conditional Heteroskedasticity (GARCH) model and Yousuf and Ng (2021) proposed two algorithm for estimating high dimensional linear time-varying parameter (TVP) model.
In this paper, we propose a time-varying STARMA (tvSTARMA) modelling, based on the wavelet expansion of coefficients.

The content of the paper is as follows. Section 2 provides basis background on locally stationary processes and wavelets. In Sections 3 and 4 the tvSTARMA models proposed and estimation procedures are presented. Some simulation studies are described in Sections 5 and 6 and an application to historical daily precipitation records is illustrated in Section 7. Some further comments are presented in Section 8.

\section{Background}

\subsection{Locally stationary process}

As mentioned in Section 1, stationarity is a basis assumption for time series analysis, however, many phenomena in the practical applications show a non stationary behavior. One way to solve this problem is to assume that the processes involved are locally stationary.

\begin{definition}
     \normalfont (Dahlahus (2000)) A sequence of Gaussian multivariate stochastic processes $X_{t,T} = (X^{(1)}_{t,T}, ..., X^{(d)}_{t,T})' \ (t = 1, ..., T)$ is called locally stationary with transfer function matrix $\textbf{A}^0$ and mean function vector $\mu$ if there exists a representation
     \begin{equation}
         X_{t,T} = \mu(\frac{t}{T}) + \int_{-\pi}^{\pi} \exp(i\lambda t) \textbf{A}^0_{t,T}(\lambda) d\xi(\lambda)
     \end{equation}
     with the following properties:
     \begin{itemize}
         \item [(i)] $\xi(\lambda)$ is a complex valued Gaussian vector process on $[-\pi, \pi]$ with $\overline{\xi_a(\lambda)} = \xi_a(-\lambda)$,  $E[\xi_a(\lambda)] =0$ and
         \begin{equation}
             E[d\xi_a(\lambda) d\xi_b(\lambda)] = \delta_{ab} \eta (\lambda+\mu)d\lambda d\mu,
         \end{equation}
         where $\eta(\lambda) = \sum_{j=-\infty}^{\infty} \delta(\lambda+2\pi j)$ is the period $2\pi$ extension of the Dirac delta function.
         \item [(ii)] There exists a constant $K$ and a $2\pi$-periodic matrix valued function \textbf{A}:$[0, 1] \times \mathbb{R} \rightarrow \mathbb{C}^{d\times d}$ with $\overline{\textbf{A}(u, \lambda)} = \textbf{A}(u, -\lambda)$ and
         \begin{equation}
             \underset{t, \lambda}{\sup} \mid \textbf{A}^0_{t,T}(\lambda)_{ab} -\textbf{A}(t/T, \lambda)_{ab} \mid \ \le \ KT^{-1}
         \end{equation}
         for all $a, b = 1,...,d$ and $T \in \mathbb{N}$. $\textbf{A}(u, \lambda)$ and $\mu(u)$ are assumed to be continuous in $\mu$.
        
         The function $f(u, \lambda):= \textbf{A}(u, \lambda) \overline{\textbf{A}(u, \lambda)}'$ is the time varying spectral density matrix of the process.
     \end{itemize}
\end{definition}

If $X_{t,T}$ is a time-varying VARMA model, it is defined by the difference equation
\begin{equation}
    \sum_{j=0}^p \Phi_j (\frac{t}{T}) \left[ X_{t-j,T} - \mu(\frac{t-j}{T}) \right] = \sum_{j=0}^q \Psi_j(\frac{t}{T}) \epsilon_{t-j},
\end{equation}
where $\epsilon_t$ are independent, identically distributed with mean zero and covariance matrix $\textbf{I}_d$ and $\Phi_0(u) \equiv \Psi_0(u) \equiv \textbf{I}_d$. For $z \in \mathbb{C}$, let $\Phi(u,z) = \sum_{j=0}^p \Phi_j(u)z^j$ and $\Psi(u,z) = \sum_{j=0}^q \Psi_j(u)z^j$, if $det(\Phi(u,z)) \ne 0$ for all $|z| \le 1+c$ with $c > 0$ uniformly in $u$ and all entries of $\Phi_j(u)$ and $\Psi_j(u)$ are continuous in $u$, then the solution of this difference equation has an infinite time-varying MA presentation, that is, the solution is locally stationary of the form (1). The time-varying spectral density of the process is 
\begin{equation}
    f(u,\lambda) = \frac{1}{2\pi} \frac{\Psi(u, e^{i\lambda})\Psi(u, e^{-i\lambda})'}{\Phi(u, e^{i\lambda}) \Phi(u, e^{-i\lambda})'}.
\end{equation}

Since a STARMA model is a special case of VARMA models, the time-varying STARMA model that will be proposed is a locally stationary process and for simplicity, in the sequel, we assume that $\mu(\cdot) = 0$.

\subsection{Wavelets}

Wavelet analysis is closely related to Fourier analysis as both enable a given function to be expressed in terms of the summation of basis functions.
The main feature of wavelet bases is that a function $f \in L^2(\mathbb{R})$ can be generated by binary dilations $2^j$ and dyadic translations $k2^{-j}$ of a scaling function $\phi(t)$ and/or of a  mother wavelet $\psi(t)$. Thus, a wavelet basis is composed of functions $\{ \phi_{j,k}(t) \cup \psi_{j,k}(t), j,k \in \mathbb{Z} \} \in L^2(\mathbb{R})$, where
\begin{equation}
    \phi_{j,k}(t) = 2^{j/2} \phi(2^j t-k),
\end{equation}
and
\begin{equation}
    \psi_{j,k}(t) = 2^{j/2} \psi(2^j t-k).
\end{equation}

One way of obtaining the scaling function is a solution of the equation
\begin{equation}
    \phi(t) = \sqrt{2} \sum_k l_k \phi(2t-k),
\end{equation}
and $\psi(t)$ is obtained from $\phi(t)$ by
\begin{equation}
    \psi(t) = \sqrt{2} \sum_k h_k \phi(2t-k),
\end{equation}

In fact, $l_k$ and $h_k$ are low-pass and high-pass filter coefficients, respectively, given by
\begin{eqnarray}
    l_k &=& \sqrt{2} \int_{-\infty}^{\infty} \phi(t)\phi(2t-k) dt, \\
    h_k &=& \sqrt{2} \int_{-\infty}^{\infty} \psi(t)\phi(2t-k) dt.
\end{eqnarray}

In this paper, the Mexican hat wavelet will be used, its mother function given by

\begin{equation}
    \psi(t) = \frac{2}{\sqrt{3}\pi^{1/4}}(1-t^2) e^{-t^2/2}.
\end{equation}

A Multi-resolution Analysis (MRA) allows us to analyze the data at different levels of resolution. The data with coarse resolution contain information about lower-frequency components and retain the main features of the original signal and the data with finer resolution retain information about the higher-frequency components. 

Let $V_j$ and $W_j$ be closed sub-spaces generated by $\{ \phi_{j,k}, k = 0,...,2^j-1 \}$ and $\{ \psi_{j,k}, k = 0,...,2^j-1 \}$, respectively, then the MRA have the following properties:
\begin{itemize}
    \item[(i)] $...\subset V_{-1} \subset V_0 \subset V_1 \subset ...$
    \item[(ii)] $L^2(\mathbb{R}) = \bigcup_j \overline{V_j}$
    \item[(iii)] $\bigcap_jV_j = \{0\}$
    \item[(iv)] $x(t) \in V_j \Leftrightarrow x(2t) \in V_{j+1}, \ \forall j$
    \item[(v)] $V_{j+1} = V_j \oplus W_j, \ W_j \perp V_j$.
\end{itemize}
The above properties imply $W_j = V_{j+1} \ominus V_j$, and then, we obtain
\begin{equation}
    L^2(\mathbb{R}) = \bigoplus_{j \in \mathbb{Z}} W_j = V_0 \oplus \bigoplus_{j \ge 0}W_j = V_{J_0} \oplus \bigoplus_{j \ge J_0}W_j
\end{equation}
for some $J_0$ integer, then, any $f(t) \in L^2(\mathbb{R})$ can be expressed as
\begin{equation}
f(t) = \sum_k c_{J_0,k}\phi_{J_0,k}(t) + \sum_{j\ge J_0}\sum_k d_{j,k} \psi_{j,k}(t),
\end{equation}
for some coarse scale $J_0$ (usually taken as zero), where
\begin{eqnarray*}
    c_{j,k} = \int_{-\infty}^{\infty} f(t)\phi_{j,k}(t)dt, \\
    d_{j,k} = \int_{-\infty}^{\infty} f(t)\psi_{j,k}(t)dt.
\end{eqnarray*}

\section{Time-varying STAR model}

\subsection{Spatial Weight Matrix}

In the construction of STARMA models, the definition of the spatial lag operator is necessary. Let $Z_i(t)$ be a random variable at location $i$ and time $t$ and $L^{(l)}$, then the spatial lag operator with spatial order $l$ is defined as
\begin{eqnarray}
    L^{(l)} Z_i(t) = \left\{\begin{array}{rl}
    Z_i(t), & \hbox{if} \ l = 0, \\
    \sum_{j=1}^n w_{ij}^{(l)} Z_j(t), & \hbox{if} \ l > 0, \end{array}\right.
    %L^{(0)} Z_i(t) &=& Z_i(t), \\
    %L^{(l)} Z_i(t) &=& \sum_{j=1}^n w_{ij}^{(l)} Z_j(t), \ l > 0,
\end{eqnarray}
where $w_{ij}^{(l)}$ are weights of order $l$ with
\begin{equation}
    \sum_{j=1}^n \omega_{ij}^{(l)} = 1.
\end{equation}
Each element of the matrix reflects the spatial relationship between two regions, $i$ and $j$ and $w^{(l)}_{ij} = 0$ when $i=j$, that is, the matrix has zeros on its main diagonal and the other elements will consist of positive numbers. In matrix form, let $\textbf{Z}(t) = [Z_1(t),...,Z_n(t)]'$ a vector of $n$ observations, 
\begin{equation}
    L^{(l)} \textbf{Z}(t) = \left\{\begin{array}{rl}
    W^{(0)} \textbf{Z}(t) = \textbf{I}_n \textbf{Z}(t), & \hbox{if} \ l = 0, \\
    W^{(l)} \textbf{Z}(t), & \hbox{if} \ l > 0, \end{array}\right.
\end{equation}
where $W^{(l)}$ is an $n \times n$ square matrix consisting of weights $w^{(l)}_{ij}$ such that the sum of items in each row is one.

\subsection{Wavelet Based tvSTAR model}

The classical STARMA ($p_{\lambda_1,...,\lambda_p}, q_{m_1,...,m_q}$) model can be defined by
\begin{equation}
    Z_i(t) = \sum_{s=1}^p \sum_{l=0}^{\lambda_s} \phi_{sl} L^{(l)} Z_i(t-s) - \sum_{s=1}^q \sum_{l=0}^{m_s} \theta_{sl} L^{(l)} \epsilon_i(t-s) + \epsilon_i(t), \ i = 1,..., n, \ t = 1,..., T,
\end{equation}
or in matrix form,
\begin{equation}
    \textbf{Z}(t) = \sum_{s=1}^p \sum_{l=0}^{\lambda_s} \phi_{sl} W^{(l)} \textbf{Z}(t-s) - \sum_{s=1}^q \sum_{l=0}^{m_s} \theta_{sl} W^{(l)} \mat{\epsilon}(t-s) + \mat{\epsilon}(t),
\end{equation}
where $p$ is the autoregressive (AR) order, $q$ is the moving average (MA) order, $\lambda_s$ is the spatial order of the $s$th AR term and $m_s$ is the spatial order of the $s$th MA term. The parameters $\phi_{sl}$ and $\theta_{sl}$ represent the parameter at time lag $s$ and space lag $l$ for the AR and MA parameter, respectively and
$\mat{\epsilon}(t)$ are independent normally distributed random errors with $E[\mat{\epsilon}(t)] = \textbf{0}$ and covariance matrix $\textbf{I}_N \sigma^2$.

Let $\textbf{Z}(\frac{t}{T}) = [Z_1(\frac{t}{T}),...,Z_n(\frac{t}{T})]'$ be a $n$-dimensional time series with $T$ observations. The time-varying STAR (tvSTAR) model is defined by
\begin{equation}
    \textbf{Z}\left( \frac{t}{T} \right) = \sum_{s=1}^p \sum_{l=0}^{\lambda_s} \phi_{sl}\left(\frac{t}{T} \right) W^{(l)} \textbf{Z}\left( \frac{t-s}{T} \right) + \mat{\epsilon}\left( \frac{t}{T} \right), \ t = 1,..., T,
\end{equation}
where $\mat{\epsilon}(\frac{t}{T})$ is an independent, identically distributed gaussian vector with mean zero, $\phi_{sl}(\frac{t}{T})$ is time-varying parameter at time lag $s$ and space lag $l$ and $W^{(l)}$ is spatial weight matrix of the order $l$. 

Wavelet bases, despite having irregular shapes, are able to perfectly reconstruct functions with linear and higher-order polynomial shapes. The idea is to expand the time-varying parameters $\phi_{sl}(\frac{t}{T})$ in wavelet expansions as
\begin{equation}
    \phi_{sl}\left( \frac{t}{T} \right) = \sum_{j=-1}^{\infty} \sum_{k = 0}^{2^j-1} \beta^{sl}_{j,k} \psi_{j,k}\left( \frac{t}{T} \right).
\end{equation}
Notice that, for simplification, we define $\psi_{-1,0}(\frac{t}{T}) = \phi_{0,0}(\frac{t}{T})$, the scaling function with $j, k =0$.

 Then, the wavelet based tvSTAR model is given by
\begin{eqnarray}
    \textbf{Z}\left( \frac{t}{T} \right) &=& \sum_{s=1}^p \sum_{l=0}^{\lambda_s} \sum_{j=-1}^{\infty} \sum_{k = 0}^{2^j-1} \beta^{sl}_{j,k} \psi_{j,k}\left( \frac{t}{T} \right) W^{(l)} \textbf{Z}\left( \frac{t-s}{T} \right) + \mat{\epsilon}\left( \frac{t}{T} \right) \nonumber \\
    &=& \sum_{s=1}^p \sum_{l=0}^{\lambda_s} \sum_{j=-1}^{J -1}\sum_{k = 0}^{2^j-1} \beta^{sl}_{j,k} \psi_{j,k}\left( \frac{t}{T} \right) W^{(l)} \textbf{Z}\left( \frac{t-s}{T} \right) \nonumber \\
    &+& \sum_{s=1}^p \sum_{l=0}^{\lambda_s} \sum_{j \ge J}^{\infty} \sum_{k = 0}^{2^j-1} \beta^{sl}_{j,k} \psi_{j,k}\left( \frac{t}{T} \right) W^{(l)} \textbf{Z}\left( \frac{t-s}{T} \right) + \mat{\epsilon}\left( \frac{t}{T} \right) \nonumber \\
    &=& \sum_{s=1}^p \sum_{l=0}^{\lambda_s} \sum_{j=-1}^{J -1}\sum_{k = 0}^{2^j-1} \beta^{sl}_{j,k} \psi_{j,k}\left( \frac{t}{T} \right) W^{(l)} \textbf{Z}\left( \frac{t-s}{T} \right) + \mat{\nu}\left( \frac{t}{T} \right),
\end{eqnarray}
where
\begin{equation}
    \mat{\nu}\left( \frac{t}{T} \right) = \sum_{s=1}^p \sum_{l=0}^{\lambda_s} \sum_{j \ge J}^{\infty} \sum_{k = 0}^{2^j-1} \beta^{sl}_{j,k} \psi_{j,k}\left( \frac{t}{T} \right) W^{(l)} \textbf{Z}\left( \frac{t-s}{T} \right) + \mat{\epsilon}\left( \frac{t}{T} \right)
\end{equation}
and $W^{(l)}$ is $n \times n$ matrix of weights with each row having sum one.
Note that $J-1$ is the the finest resolution level such that $2^{J-1} \le \sqrt{T} \le 2^J$, and then, $\sqrt{n}2^J/2 \le \sqrt{nT} \le \sqrt{n}2^J$ with $n \ge 2$.

The linear form of (22) can be written as
\begin{equation}
    \textbf{Z} = \mat{\Psi}\mat{\beta} + \mat{\nu},
\end{equation}
where 
\begin{eqnarray}
    \textbf{Z} &=& \left[ \hbox{vec} \left( \textbf{Z}\left( \frac{p+1}{T} \right)\right)',..., \hbox{vec} \left(\textbf{Z}\left( \frac{T}{T} \right) \right)'\right]', \\
    \mat{\nu} &=& \left[ \hbox{vec} \left( \mat{\nu}\left( \frac{p+1}{T} \right)\right)',..., \hbox{vec}\left(\mat{\nu}\left( \frac{T}{T} \right) \right)'\right]', \\
    \mat{\Psi} &=& \left[ \Psi_{-1,0}, \Psi_{0,0},...,\Psi_{J-1, 2^{J-1}-1} \right]
\end{eqnarray}
and
\begin{equation}
    \Psi_{j,k} = 
    \begin{bmatrix}
        \psi_{j,k}\left( \frac{p+1}{T} \right)\sum_{m=1}^n w^{(0)}_{1,m} Z_m \left( \frac{p}{T} \right) & \dots & \psi_{j,k}\left( \frac{p+1}{T} \right)\sum_{m=1}^n w^{(\lambda_p)}_{1,m} Z_m \left( \frac{1}{T} \right)\\
        \vdots & & \vdots \\
        \psi_{j,k}\left( \frac{T}{T} \right)\sum_{m=1}^n w^{(0)}_{1,m} Z_m \left( \frac{T-1}{T} \right) & \dots & \psi_{j,k}\left( \frac{T}{T} \right)\sum_{m=1}^n w^{(\lambda_p)}_{1,m} Z_m \left( \frac{T-p}{T} \right)\\
        \psi_{j,k}\left( \frac{p+1}{T} \right)\sum_{m=1}^n w^{(0)}_{2,m} Z_m \left( \frac{p}{T} \right) & \dots & \psi_{j,k}\left( \frac{p+1}{T} \right)\sum_{m=1}^n w^{(\lambda_p)}_{2,m} Z_m \left( \frac{1}{T} \right)\\
        \vdots & & \vdots \\
        \psi_{j,k}\left( \frac{T}{T} \right)\sum_{m=1}^n w^{(0)}_{n,m} Z_m \left( \frac{T-1}{T} \right) & \dots & \psi_{j,k}\left( \frac{T}{T} \right)\sum_{m=1}^n w^{(\lambda_p)}_{n,m} Z_m \left( \frac{T-p}{T} \right)\\
    \end{bmatrix},
\end{equation}
for $j = -1, ...,J-1, \ k = 0,...,2^j-1$, where $\mat{\beta}$ is a $s_a(2^J) \times 1$ vector containing the wavelet coefficients, with $s_a = \sum_{s=1}^p(1+ \lambda_s)$.

\subsection{Estimation}

The number of parameters depends on the spatial order, $\lambda_s$, for $s = 1,..,p$ and the resolution level $J$. For really large datasets, there are many parameters to be estimated, the method of least squares is a computationally efficiency estimation method in this case.

Assume that the matrix $\mat{\Psi}'\mat{\Psi}$ is positive definite, the least squares estimator of $\mat{\beta}$ is given by
\begin{equation}
    \hat{\mat{\beta}} = (\mat{\Psi}'\mat{\Psi})^{-1} \mat{\Psi}' \textbf{Z}.
\end{equation}

We consider the following assumptions:
\begin{itemize}
    \item [A.1] (Dahlhaus et al. (1999)) The functions $\phi_{kl}$ are real, bounded and belong to the following set of functions:
    \begin{equation}
        \mathscr{F} = \left\{ f(x): f(x) = \sum_{j=-1}^{\infty}  \sum_{k=0}^{\infty} \beta_{j,k} \psi_{j,k}(x) \middle\vert\ \norm{\beta_{..}}_{d,p,q} < \infty \right\},
    \end{equation}
    where
    \begin{equation}
        \norm{\beta_{..}}_{d,p,q} = \left[ \sum_{j \ge -1} \left( 2^{jup} \sum_{k=0}^{2^j-1}|\beta_{j,k}|^p \right)^{q/p} \right]^{1/q},
    \end{equation}
    $u=d+1/2-1/\Tilde{p} > 1$ with $\Tilde{p} = \hbox{min}\{p,2\}$. Note that $d$ is the degree of smoothness and $1 < p,q \le \infty$ specify the norm in which smoothness is measured.
    \item [A.2] The functions $\phi(t)$ and $\psi(t)$ belong to $C^r[0,1]$, with degree of regularity $r > d$ and they have compact support. In addition, $\int \phi(t) dt = 1$ and $\int \psi(t)^k dt = 0$ for $0 \le k \le r$.
    \item [A.3] There exists some $\gamma \ge 0$ with $|\hbox{cum}_n(\nu_t)| \le A^n(n!)^{1+\gamma}$ for all $n, t$, where $A$ is a positive constant and $\nu_t = \nu(\frac{t}{T})$. Here, $\hbox{cum}_n(\nu_t)$ denotes the cumulant of order $n$ of $\nu_t$
    \item [A.4] $\textbf{Z}(\frac{t}{T})$ are locally stationary processes.
    
\end{itemize}

\begin{prop}
    \normalfont    
    Suppose the assumptions A.1-A.4 hold, then
    \begin{itemize}
        \item [(i.)] $E(\hat{\mat{\beta}}) = O((nT)^{-1/2})$.
        \item [(ii.)] $E[ (\hat{\mat{\beta}} - \mat{\beta})(\hat{\mat{\beta}} - \mat{\beta})' ] = O((nT)^{-1})$.
    \end{itemize}
\end{prop}

\textbf{Proof.} See the Appendix.

\begin{prop}
    \normalfont
    Suppose the assumptions A.1-A.4 hold, then we have
    \begin{equation}
        \sqrt{N}\textbf{H}(\hat{\mat{\beta}} - \mat{\beta}) \mathop\to^D \mathcal{N}_K(\textbf{0},\mat{\Gamma}),
    \end{equation}
    where $D$ means convergence in distribution and covariance matrix 
    \begin{equation}
        \mat{\Gamma} = \lim_{N \to \infty} N \textbf{H} E[\mat{\Psi}'\mat{\Psi}]^{-1} \textbf{H}',
    \end{equation}
    where $N = n(T-p)$ and \textbf{H} is a matrix with $K$ rows.
\end{prop}

\textbf{Proof.} See the Appendix.

\section{Wavelet based time-varying STARMA model}

\subsection{The model}

Let $\textbf{Z}\left( \frac{t}{T} \right) = [Z_1(\frac{t}{T}), Z_2(\frac{t}{T}), ..., Z_n(\frac{t}{T})]'$ be a multivariate locally stationary of $n$-dimensional time-varying STARMA (tvSTARMA) model, defined by
\begin{equation}
    \textbf{Z}\left( \frac{t}{T} \right) = \sum_{s=1}^p \sum_{l=0}^{\lambda_s} \phi_{sl}\left(\frac{t}{T} \right) W^{(l)}\textbf{Z} \left( \frac{t-s}{T} \right) - \sum_{s=1}^q \sum_{l=0}^{m_s} \theta_{sl}\left(\frac{t}{T}\right) W^{(l)} \mat{\epsilon}\left( \frac{t-s}{T} \right) + \mat{\epsilon}\left( \frac{t}{T} \right), \ t = 1,..., T,
\end{equation}
where $\mat{\epsilon}(\frac{t}{T})$ is independent, identically distributed gaussian vector with mean zero, $\phi_{sl}(\frac{t}{T})$ and $\theta_{sl}(\frac{t}{T})$ are time-varying parameters at time lag $s$ and space lag $l$ and $W^{(l)}$ is a spatial weight matrix of the order $l$.

The time-varying parameters $\phi_{sl}(\frac{t}{T})$ and $\theta_{sl}(\frac{t}{T})$ can be written in wavelet expansions as
\begin{eqnarray}
    \phi_{sl}\left( \frac{t}{T} \right) = \sum_{j=-1}^{\infty} \sum_{k = 0}^{2^j-1} a^{sl}_{j,k} \psi_{j,k}\left( \frac{t}{T} \right), \\
    \theta_{sl}\left( \frac{t}{T} \right) = \sum_{j=-1}^{\infty} \sum_{k = 0}^{2^j-1} b^{sl}_{j,k} \psi_{j,k}\left( \frac{t}{T} \right).
\end{eqnarray}

Replacing equations (35), (36) into (34), the wavelet based tvSTARMA model is given by

\begin{eqnarray}  
    \textbf{Z}\left( \frac{t}{T} \right) &=& \sum_{s=1}^p \sum_{l=0}^{\lambda_s} \sum_{j=-1}^{\infty} \sum_{k = 0}^{2^j-1} a^{sl}_{j,k} \psi_{j,k}\left( \frac{t}{T} \right) W^{(l)} \textbf{Z}\left( \frac{t-s}{T} \right) \nonumber \\
    &-& \sum_{s=1}^q \sum_{l=0}^{m_s} \sum_{j=-1}^{\infty} \sum_{k = 0}^{2^j-1} b^{sl}_{j,k} \psi_{j,k}\left( \frac{t}{T} \right) W^{(l)} \mat{\epsilon}\left( \frac{t-s}{T} \right) + \mat{\epsilon}\left( \frac{t}{T} \right) \nonumber \\
    &=& \sum_{s=1}^p \sum_{l=0}^{\lambda_s} \sum_{j=-1}^{J -1}\sum_{k = 0}^{2^j-1} a^{sl}_{j,k} \psi_{j,k}\left( \frac{t}{T} \right) W^{(l)} \textbf{Z}\left( \frac{t-s}{T} \right) \nonumber \\
    &-& \sum_{s=1}^q \sum_{l=0}^{m_s} \sum_{j=-1}^{J-1} \sum_{k = 0}^{2^j-1} b^{sl}_{j,k} \psi_{j,k}\left( \frac{t}{T} \right) W^{(l)} \mat{\epsilon}\left( \frac{t-s}{T} \right) \nonumber \\
    &+& \sum_{s=1}^p \sum_{l=0}^{\lambda_s} \sum_{j \ge J}^{\infty} \sum_{k = 0}^{2^j-1} a^{sl}_{j,k} \psi_{j,k}\left( \frac{t}{T} \right) W^{(l)} \textbf{Z}\left( \frac{t-s}{T} \right) \nonumber \\
    &-& \sum_{s=1}^q \sum_{l=0}^{m_s} \sum_{j \ge J}^{\infty} \sum_{k = 0}^{2^j-1} b^{sl}_{j,k} \psi_{j,k}\left( \frac{t}{T} \right) W^{(l)} \mat{\epsilon}\left( \frac{t-s}{T} \right) + \mat{\epsilon}\left( \frac{t}{T} \right) \nonumber \\
    &=& \sum_{s=1}^p \sum_{l=0}^{\lambda_s} \sum_{j=-1}^{J -1}\sum_{k = 0}^{2^j-1} a^{sl}_{j,k} \psi_{j,k}\left( \frac{t}{T} \right) W^{(l)} \textbf{Z}\left( \frac{t-s}{T} \right) \nonumber \\
    &-& \sum_{s=1}^q \sum_{l=0}^{m_s} \sum_{j=-1}^{J-1} \sum_{k = 0}^{2^j-1} b^{sl}_{j,k} \psi_{j,k}\left( \frac{t}{T} \right) W^{(l)} \mat{\epsilon}\left( \frac{t-s}{T} \right) + \mat{\nu}\left( \frac{t}{T} \right),
\end{eqnarray}
where
\begin{eqnarray}
    \mat{\nu}\left( \frac{t}{T} \right) &=& \sum_{s=1}^p \sum_{l=0}^{\lambda_s} \sum_{j \ge J}^{\infty} \sum_{k = 0}^{2^j-1} a^{sl}_{j,k} \psi_{j,k}\left( \frac{t}{T} \right) W^{(l)} \textbf{Z}\left( \frac{t-s}{T} \right) \nonumber \\
    &-& \sum_{s=1}^q \sum_{l=0}^{m_s} \sum_{j \ge J}^{\infty} \sum_{k = 0}^{2^j-1} b^{sl}_{j,k} \psi_{j,k}\left( \frac{t}{T} \right) W^{(l)} \mat{\epsilon}\left( \frac{t-s}{T} \right)  + \mat{\epsilon}\left( \frac{t}{T} \right).
\end{eqnarray}

\subsection{Estimation}

As in the case of the STARMA model, linear and non-linear estimators can be used for the tvSTARMA model's estimation. 
%Considering the natural of the analysed variables and the non-linear spatial dependence, the non-linear estimation may be more efficient and robust. 
Maximum likelihood estimation is a widely used technique, but it is computationally expensive when there is a large number of parameters to be estimated and can be sensitive to the choice of starting values and then the optimization algorithms may converge to a local minimum or even not converge. 

An alternative is the Kalman filter (Kalman (1960)), associated with the state space model %the filter is named after Rudolf E. Kálmán, who published the famous paper Kalman (1960) and is a concept much applied in time series analysis. The Kalman filter can be used in discrete linear dynamic system
\begin{eqnarray}
    \textbf{x}_{t+1} &=& F_t \textbf{x}_t + \Gamma_t \textbf{w}_{t+1}, \\
    \textbf{z}_t &=& M_t \textbf{x}_t + \textbf{v}_t,
\end{eqnarray}
where (39) is the state equation and (40) is the observation equation of the system, $\textbf{x}_t$ is the state variable at time $t$, $\textbf{z}_t$ is the observation at $t$, $F_t, \Gamma_t$ and $M_t$ are state transition matrix, control input matrix and observation matrix, respectively. The vectors $\textbf{v}_t$ and $\textbf{w}_t$ are errors assumed to be Gaussian.
The algorithm provides recursive estimation formulas to estimate the state of a process, in the sense of minimizing the squared error.

For adaptive parameter estimation in the tvSTARMA model by Kalman filter, first, we rewrite (37) in linear form
\begin{equation}
     \textbf{Z}\left( \frac{t}{T} \right) = \textbf{Y}\left( \frac{t}{T} \right) \textbf{c} + \mat{\nu}\left( \frac{t}{T} \right),
\end{equation}

where $\mat{\nu}\left( \frac{t}{T} \right)$ is assume to have a multivariate normal distribution with mean zero and covariance matrix $\Sigma$,

\begin{align*}
    s_a &= \sum_{s=1}^p(1+ \lambda_s), \\
    s_m &= \sum_{s=1}^q(1+ m_s), \\
    \textbf{c}_{2^J[s_a+s_m] \times 1} &= 
    \begin{bmatrix}
        \textbf{a} \\
        \textbf{b}
    \end{bmatrix}, \\
    \textbf{a}'_{2^J [s_a] \times 1} &= [a^{10}_{-1,0} \ a^{10}_{0,0} \ ... \ a^{10}_{J-1,2^{J-1}-1} \ ... \ a^{p\lambda_p}_{J-1,2^{J-1}-1}], \\
    \textbf{b}'_{2^J [s_m] \times 1} &= [b^{10}_{-1,0} \ b^{10}_{0,0} \ ... \ b^{10}_{J-1,2^{J-1}-1} \ ... \ b^{qm_q}_{J-1,2^{J-1}-1}], \\
    \textbf{Y}\left( \frac{t}{T} \right)_{n \times 2^J[s_a+s_m]} &= \left[ \hbox{diag}_n \otimes \textbf{1}_{1 \times [s_a+s_m]} \textbf{W} \textbf{D}\left( \frac{t}{T} \right) \right] \otimes \Psi\left( \frac{t}{T} \right), \ \hbox{where}, \\
    \Psi\left( \frac{t}{T} \right)'_{2^J \times 1} &= \left[ \psi_{-1,0}\left( \frac{t}{T} \right) \ \psi_{0,0}\left( \frac{t}{T} \right) \ ... \ \psi_{J-1,2^{J-1}-1}\left( \frac{t}{T} \right) \right], \\
    \textbf{W}_{n[s_a+s_m] \times n[s_a+s_m] } &=
    \mqty[\dmat{\textbf{W}^{ar},\textbf{W}^{ma}}], \\
    \textbf{W}^{ar}_{n \times n[s_a]} &= \mqty[\dmat{W^{(0)},\ddots, W^{(\lambda_1)}, W^{(0)}, \ddots, W^{(\lambda_2)},\ddots, W^{(\lambda_p)}}], \\
    \textbf{W}^{ma}_{n \times n[s_m]} &= \mqty[\dmat{W^{(0)},\ddots, W^{(m_1)}, W^{(0)}, \ddots, W^{(m_2)},\ddots, W^{(m_q)}}], \\
    \textbf{D}\left( \frac{t}{T} \right)_{n[s_a+s_m] \times [s_a+s_m]} &= 
    \begin{bmatrix}
        \textbf{X}\left( \frac{t}{T} \right) &  \\
          & \textbf{E}\left( \frac{t}{T} \right)
    \end{bmatrix}, \\
    \textbf{X}\left( \frac{t}{T} \right)_{n[s_a] \times s_a} &=
    \mqty[\dmat{\hbox{diag}_{(1+\lambda_1)} \otimes \textbf{Z}(\frac{t-1}{T}),\hbox{diag}_{(1+\lambda_2)} \otimes \textbf{Z}(\frac{t-2}{T}), \ddots, \hbox{diag}_{(1+\lambda_p)} \otimes \textbf{Z}(\frac{t-p}{T})}], \\
    \textbf{E}\left( \frac{t}{T} \right)_{n[s_m] \times s_m} &= 
    \mqty[\dmat{\hbox{diag}_{(1+m_1)} \otimes \mat{\epsilon}(\frac{t-1}{T}),\hbox{diag}_{(1+m_2)} \otimes \mat{\epsilon}(\frac{t-2}{T}), \ddots, \hbox{diag}_{(1+m_q)} \otimes \mat{\epsilon}(\frac{t-q}{T})}].
\end{align*}

According to Cipra and Motykova (1987), the state space representation (39) and (40) for the estimation of the parameters $\textbf{c}$ can be written as
\begin{eqnarray}
    \textbf{c}_{t+1} &=& \textbf{c}_t, \\
    \textbf{z}_t &=& \textbf{y}_t \textbf{c}_t + \mat{\epsilon}_t,
\end{eqnarray}
where $F_t = \textbf{I}_{2^J[s_a+s_m]}$, $\Gamma_t = \textbf{0}_{2^J[s_a+s_m]}$, $\textbf{z}_t = \textbf{Z}(\frac{t}{T})$, $M_t = \textbf{y}_t = \textbf{Y}(\frac{t}{T})$, $\textbf{v}_t = \mat{\nu}_t = \mat{\nu}(\frac{t}{T})$. Then, the recursive equations have the following form:
\begin{eqnarray}
    \hat{\textbf{c}}_{t+1} &=& \hat{\textbf{c}}_t + \textbf{P}_{t+1} \textbf{y}'_{t+1} \hat{\mat{\Sigma}}_t^{-1}(\textbf{z}_{t+1} - \textbf{y}_{t+1} \hat{\textbf{c}}_t), \\
    \textbf{P}_{t+1} &=& \textbf{P}_t - \textbf{P}_t \textbf{y}'_{t+1}(\textbf{y}_{t+1}\textbf{P}_t\textbf{y}'_{t+1} + \hat{\mat{\Sigma}}_t)^{-1}\textbf{y}'_{t+1}\textbf{P}_t, \\
    \hat{\mat{\Sigma}}_{t+1} &=& \frac{1}{t+1-(s_a+s_m)}\{ [t-(s_a+s_m)]\hat{\mat{\Sigma}}_t + (\textbf{z}_{t+1} - \textbf{y}_{t+1} \hat{\textbf{c}}_{t+1})(\textbf{z}_{t+1} - \textbf{y}_{t+1} \hat{\textbf{c}}_{t+1})' \}, \\
    \hat{\mat{\nu}}_{t+1} &=& \textbf{z}_{t+1} - \textbf{y}_{t+1} \hat{\textbf{c}}_{t+1}.
\end{eqnarray}

Note that if there is no a priori information on the parameters, the initial values of the estimates at time $t_0$ can be chosen as
\begin{equation}
    \hat{\textbf{c}}_{t_0} = \textbf{0}_{2^J[s_a+s_m] \times 1}, \ \textbf{P}_{t_0} = \textbf{I}_{2^J[s_a+s_m] \times 1}, \ \hat{\mat{\Sigma}}_{t_0} = h\textbf{I}_{n \times n},
\end{equation}
where $h$ is a small positive constant.

\section{Simulations}

This section presents some simulation examples in order to evaluate the performance of the proposed estimation procedure. 

\subsection{Simulation Procedure}

The simulations consist of the following steps:
\begin{itemize}
    \item[[1.]] Let $n$ = 15 sample locations generated (as shown in Figure 1), and then, $M$ = 1000 experiments of $n$ time series with length $T$ = 1024 are simulated.
    \item[[2.]] Each sample data \textbf{z} = [$\textbf{z}(\frac{1}{T}), ..., \textbf{z}(\frac{T}{T})$] is simulated from a tvSTAR($1_1$) process 
    \begin{equation}
        \textbf{z}\left( \frac{t}{T} \right) = \phi_{10}(t)W^{(0)}\textbf{z}\left( \frac{t-1}{T} \right) + \phi_{11}(t)W^{(1)}\textbf{z}\left( \frac{t-1}{T} \right)+\mat{\epsilon}\left( \frac{t}{T} \right)
    \end{equation}
    or a tvSTARMA($1_1, 1_1$) process
    \begin{eqnarray}
        \textbf{z}\left( \frac{t}{T} \right) &=& \phi_{10}(t)W^{(0)}\textbf{z}\left( \frac{t-1}{T} \right) + \phi_{11}(t)W^{(1)}\textbf{z}\left( \frac{t-1}{T} \right) \nonumber \\
        &+& \theta_{10}(t)W^{(0)}\mat{\epsilon}\left( \frac{t-1}{T} \right) + \theta_{11}(t)W^{(1)}\mat{\epsilon}\left( \frac{t-1}{T} \right)+\mat{\epsilon}\left( \frac{t}{T} \right),
    \end{eqnarray}
    where
    $\mat{\epsilon}(t)$ is a multivariate normal process with mean zero and covariance matrix $\textbf{I}_{nT}\sigma^2$, with $\sigma^2 = 1$. The elements of the spatial weight matrix $W$ are defined by
    \begin{equation}
        w_{ij} = \frac{d^{-0.5}_{ij}}{\sum_{k \ne i} d^{-0.5}_{ik}},
    \end{equation}
    $d_{ij}$ is the orthodromic distance (great-circle or spherical distance) between two locals $\textbf{x}_i$ and $\textbf{x}_j$. It is the shortest
    distance between two points on the surface of a sphere. Let $\textbf{x}_i = (\xi_i,\kappa_i)$ and $\textbf{x}_j = (\xi_j,\kappa_j)$ be the geographical locations, where $\xi_{\eta}$ and $\kappa_{\eta}, \eta = i,j,$ are the latitude and longitude, respectively. The distance $d_{ij}$ is given by 
    \begin{equation}
        d_{ij} = R \sigma,
    \end{equation}
    where $R$ is radius of the Earth and
    \begin{equation}
        \sigma = \arccos[\sin \xi_i \sin \xi_j + \cos \xi_i \cos \xi_j \cos (|\kappa_i - \kappa_j|)].
    \end{equation}
    \item[[3.]] Calculate the estimates of $\phi_{10}(t), \phi_{11}(t)$ and/or $\theta_{10}(t), \theta_{11}(t)$ using the estimation procedure described in Sections 3.3 and 4.2.
    \item[[4.]] Calculate mean squared error (MSE) of the predictors $\hat{\textbf{z}}$,
    \begin{equation}
    \hbox{MSE} = \frac{1}{nT} \sum_{i=1}^n \sum_{t= 1}^T [z_i\left( \frac{t}{T} \right) - \hat{z}_i\left( \frac{t}{T} \right)]^2,
    \end{equation}
    where $\hat{z}_i\left( \frac{t}{T} \right)$ is the estimator of $z_i\left( \frac{t}{T} \right)$, the observation at location $i$ and time $t$.
\end{itemize}

\subsection{Formulation of time-varying parameter}\mbox{}
Two groups of parameters will be used to generate the sample data, namely
\begin{itemize}
    \item[(i)] Group 1:
        \begin{eqnarray}
            \phi_{10}(t) &=& 0.5 - \frac{ \sin(\frac{2\pi t}{T})}{4}, \\ \phi_{11}(t) &=& -0.5 - \frac{ \cos(\frac{2\pi t}{T})}{4};
        \end{eqnarray}
    \item[(ii)] Group 2:
        \begin{eqnarray}
            \phi_{10}(t) &=& 0.5\left(1-\frac{t}{T}\right)^2, \\
            \phi_{11}(t) &=& -0.5\left(1-\frac{t}{T}\right)^2, \\
            \theta_{10}(t) &=& 0.5\left(\frac{t}{T}\right)^2, \\
            \theta_{11}(t) &=& -0.5\left(\frac{t}{T}\right)^2.
        \end{eqnarray}
\end{itemize}

The series of one location of the sampled tvSTAR($1_1$) and tvSTARMA($1_1, 1_1$) processes simulated are presented in Figure 2.

\begin{figure}[h]
    \centering
    \includegraphics[scale=0.55]{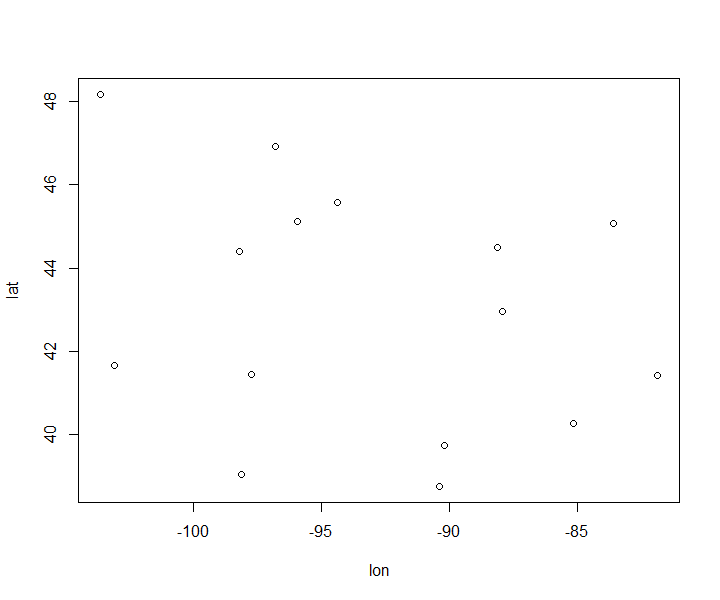}
    \caption{Simulated locations.}
    \label{fig:my_label}
\end{figure}

\begin{figure}[H]
    \centering
    \subfloat[Group 1]{\includegraphics[width=0.4\textwidth, keepaspectratio]{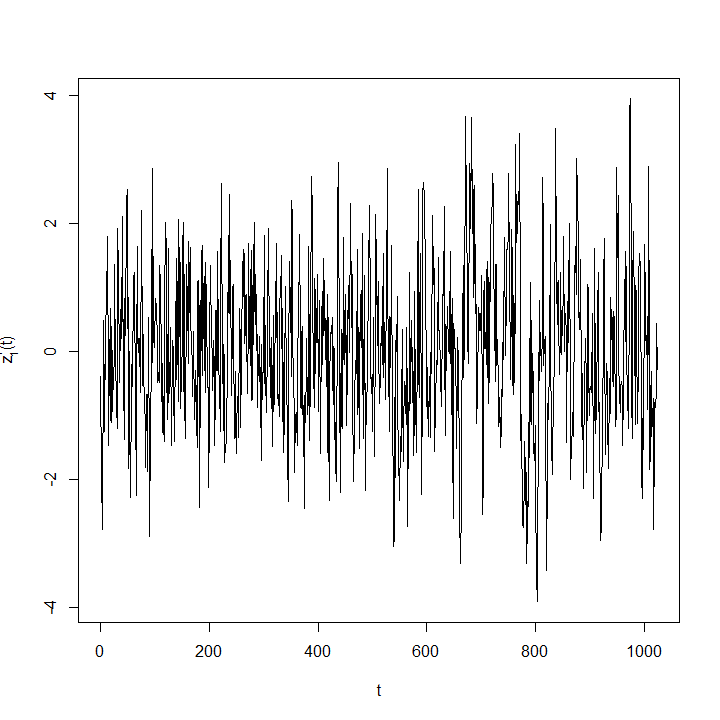}\label{fig:taba}}
    \subfloat[Group 2]{\includegraphics[width=0.4\textwidth, keepaspectratio]{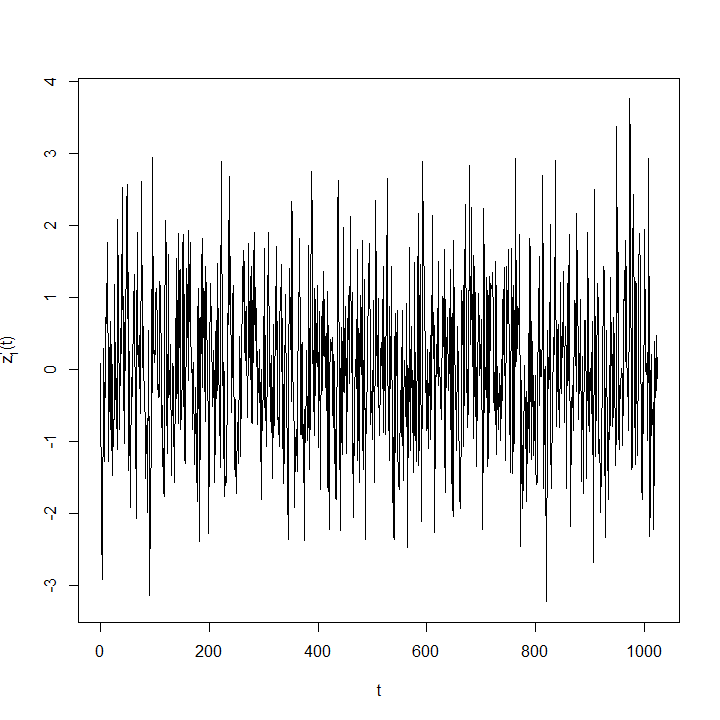}\label{fig:tabb}}
    \caption{Simulated series for one location.}
    \label{fig:my_label}
\end{figure}

\subsection{Results}

Since the true parameters are smooth functions, the Mexican hat wavelet was chosen.
Note that all estimates were calculated with $J=2$.

Figure 3 shows the boxplots of MSEs of the estimates of \textbf{z}.
We can see that the first three quartiles of the MSEs of
both groups are about the same, both groups showing outliers and Group 2 having a lower variability.

Figures 4 and 5 show the comparison of the true parameters against the average of estimates over 1000 experiments
of Group 1 and 2, respectively. We see a good fit in both cases. Note that the estimates of Group 1 are better maybe
because we have a tvSTAR model, while for Group 2, MA terms are incorporated. In conclusion, for both groups,    we have a very good performance of the proposed estimation procedure.

\begin{figure}[h!]
    \centering
    \includegraphics[width=11.5cm]{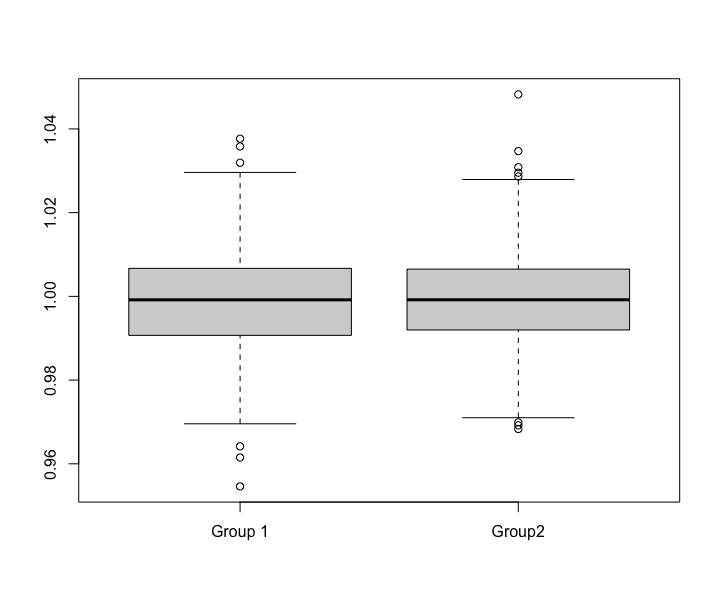}
    \caption{Boxplots of MSEs of 1000 experiments.}
    \label{fig:my_label}
\end{figure}

\begin{figure}[h!]
  \centering
  \begin{tabular}{cc}
    \includegraphics[width=8cm]{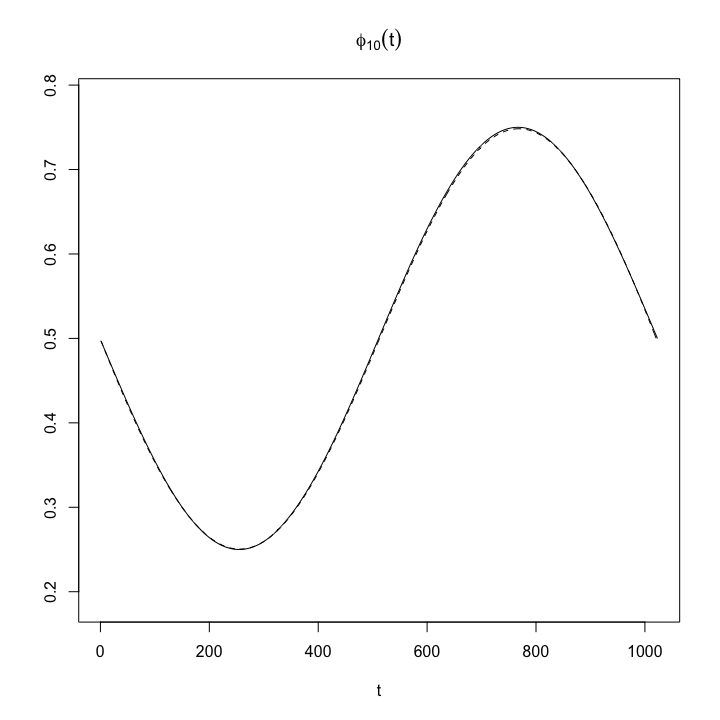}
    &
    \includegraphics[width=8cm]{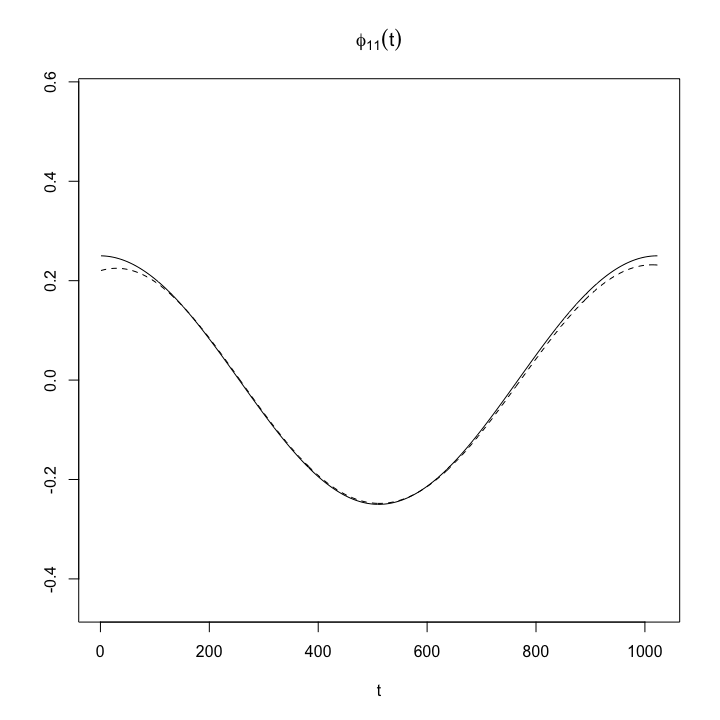}
  \end{tabular}
  \caption{Comparison of the true parameters (solid line) versus the averages of estimates obtained by Mexican hat wavelet (dotted line) of Group 1.}
\end{figure}

\begin{figure}[htb!]
  \centering
  \begin{tabular}{ll}
   \includegraphics[width=8cm]{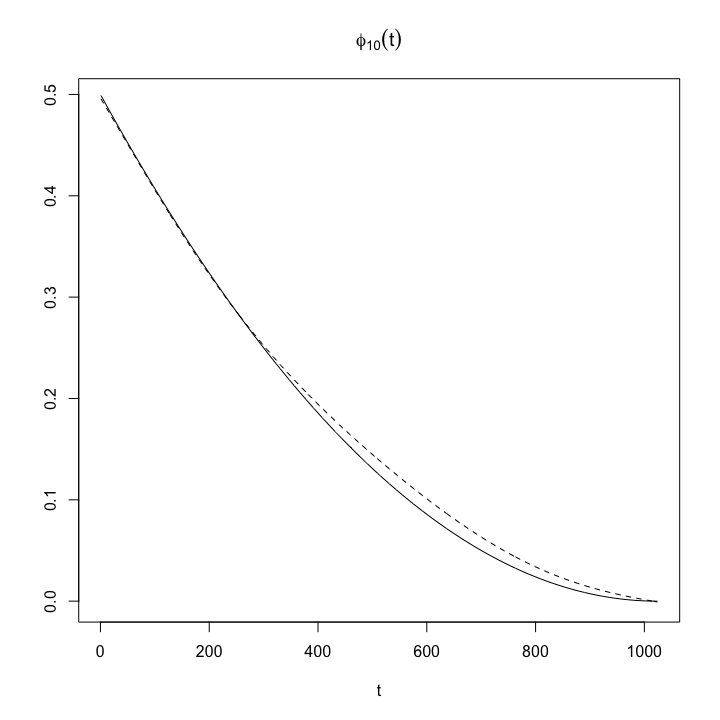}
    & \includegraphics[width=8cm]{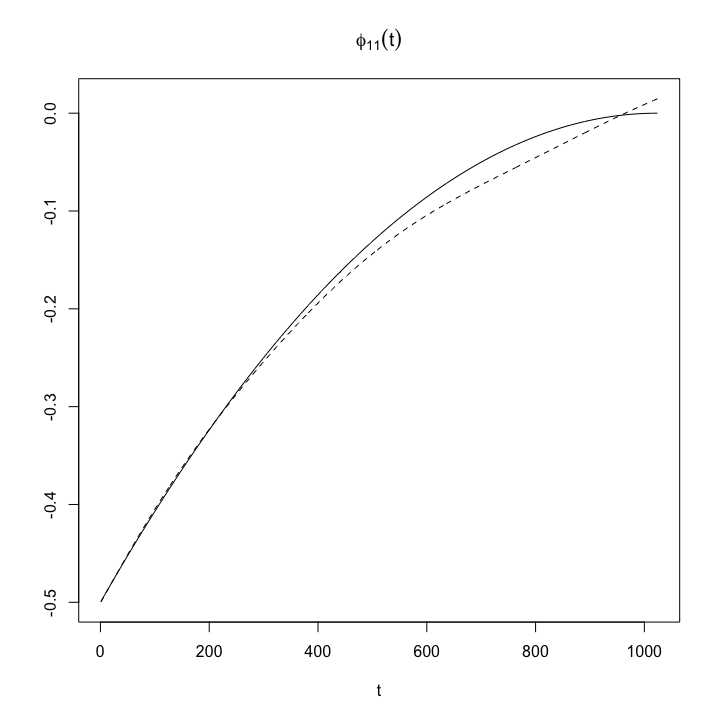}
    \\
    \includegraphics[width=8cm]{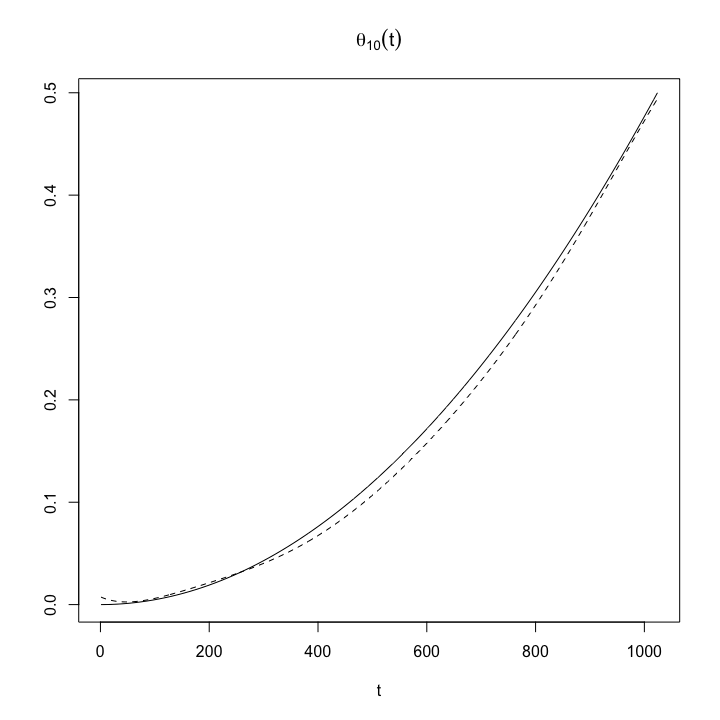}
    & \includegraphics[width=8cm]{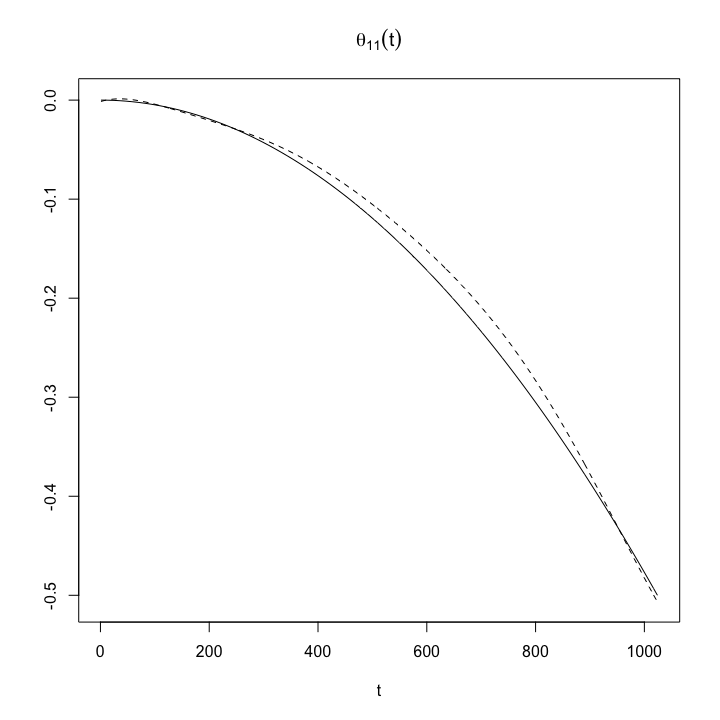}
  \end{tabular}
  \caption{Comparison of the true parameters (solid line) versus the averages of estimates obtained by Mexican hat wavelet (dotted line) of Group 2.}
\end{figure}

\section{Comparisons of spatial weight matrices via simulation}

Inspired by Jin (2017), this section presents some simulation examples in order to compare different types of spatial weight matrices. As mentioned in Section 3.1, in the tvSTARMA models, the spatial weight matrix is also a main feature.
There are many choices for the distances of the matrix $W$, such as inverse, negative exponential, $k$-nearest neighbors, etc. The inverse weight and negative exponential weight distances are selected in our simulations. They are defined by
\begin{equation}
    w^{di}_{ij} = \left\{\begin{array}{rl}
    \frac{d^{-\alpha}_{ij}}{\sum_{k \ne i} d^{-\alpha}_{ik}}, & i \ne j, \\
    0, & i = j \end{array}\right.
\end{equation}
and
\begin{equation}
    w^{ne}_{ij} = \left\{\begin{array}{rl}
    \frac{\exp{-\alpha d_{ij}}}{\sum_{k \ne i} \exp{-\alpha d_{ik}}}, & i \ne j, \\
    0, & i = j, \end{array}\right.
\end{equation}
where, $d_{ij}$ is great-circle distance and both depend on the parameter $\alpha$, a positive number.

\subsection{Simulation procedure}

The simulations consist of the following steps:
\begin{itemize}
    \item[[1.]] We consider $n$ = 15 sample locations generated as in Figure 1, and then, $M$ = 500 experiments of $n$ time series with length $T$ = 1024 are simulated.
    \item[[2.]] The sample data \textbf{z} = [$\textbf{z}(\frac{1}{T}), ..., \textbf{z}(\frac{T}{T})$] are simulated from a Gaussian random field with mean function $\mu(\textbf{z}) = \textbf{0}$ and covariance function
    \begin{equation}        
        \hbox{Cov}(z_{\textbf{x}+\textbf{h}}(t+\tau), z_{\textbf{x}}(t)) = \frac{\sigma^2}{\beta(|\tau|)^{d/2}} \omega \left\{ \frac{||\textbf{h}||}{\sqrt{\beta(|\tau|)}} \right\} + \sigma_{\epsilon}^2 \mathbb{I}_{(z_{\textbf{x}+\textbf{h}}(t+\tau) = z_{\textbf{x}}(t))},(\textbf{h}; \tau) \in R^d \times R,
     \end{equation}   
     where \textbf{h} and $\tau$ are spatial and temporal shifts, respectively, $\sigma^2 = \hbox{Var}(z_\textbf{x}(t))$, $d = 2$  and $\sigma_{\epsilon}^2 = 0.05$; $\omega(\cdot) \ge 0$ is a completely monotone function and $\beta(\cdot)$ is a positive function with a completely monotone derivative, given by
    \begin{eqnarray}
        \beta(f) &=& (f^{\zeta}+1)^{\delta/\zeta}, \\
        \omega(f) &=& \exp{-f/\gamma},
    \end{eqnarray}
    where $\zeta = 1$, $\delta = 0.5,1,1.5$ and $\gamma = 0.25,0.5,1$ are selected.
    \item[[3.]] Fit tvSTAR($1_1$) and tvSTARMA($1_1,1_1$) models using the estimation procedure described in Sections 3.3 and 4.2 with different spatial weight matrices, where we set $\alpha$ = 0.5, 1 for the inverse weight distance and $\alpha$ = 0.5, 1, 2 for the negative exponential weight distance.
    \item[[4.]] Calculate mean squared error (MSE) of the predictors $\hat{\textbf{z}}$,
    \begin{equation}
    \hbox{MSE} = \frac{1}{MnT} \sum_{m=1}^M \sum_{i=1}^n \sum_{t= 1}^T [z^m_i\left( \frac{t}{T} \right) - \hat{z}^m_i\left( \frac{t}{T} \right)]^2,
    \end{equation}
    where $\hat{z}^m_i\left( \frac{t}{T} \right)$ is the estimator of $z^m_i\left( \frac{t}{T} \right)$, the observation at location $i$ and time $t$ of the $m$th experiment.
\end{itemize}

\subsection{Results}

From Tables 1-3, we can say that: (i) for any values of the parameters, the MSEs decreases slightly for larger J; (ii) for any values of the parameters, the MSEs decreases as $\alpha$ increases; (iii) the MSEs are smaller for $\delta$ = 0.5 and any $\gamma$; (iv) the MSEs are smaller for the tvSTARMA($1_1, 1_1$) models; (v) matrices with inverse distance and $\alpha$ = 1 present the smaller MSEs for almost all cases.
For these situations, the effects of the spatial weights matrices are similar.

\begin{table}[htb]
    \centering
    \caption{MSE of different datasets from fitted tvSTAR($1_1$) model with Mexican hat wavelet and $J=2$.}
    \begin{tabular}{cccccc}
    \toprule
    %\multicolumn{6}{c}{Haar}\\ 
    %\bottomrule
    \multirow{2}*{Spatial weight matrix} & \multicolumn{2}{c}{$w^{di}_{ij}$} & \multicolumn{3}{c}{$w^{ne}_{ij}$}\\ \cline{2-6}
     &  $\alpha=0.5$ & $\alpha=1$ & $\alpha=0.5$ & $\alpha=1$ & $\alpha=2$ \\\hline
    $\gamma = 0.25$ $\delta = 0.5$ & 0.5720284 & 0.5720198 & 0.5720292 & 0.5720263 & \textbf{0.5720187} \\   
    $\gamma = 0.25$ $\delta = 1$ & 0.8099595 & \textbf{0.8099531}  & 0.8099612 & 0.8099605 & 0.8099591 \\ 
    $\gamma = 0.25$ $\delta = 1.5$ & 0.9289999 & \textbf{0.9289853}  & 0.9290054 & 0.9290036 & 0.9290000 \\
    $\gamma = 0.5$ $\delta = 0.5$ & 0.5719711 & \textbf{0.5718976} & 0.5719838 & 0.5719739 & 0.5719395 \\
    $\gamma = 0.5$ $\delta = 1$ & 0.8097927 & \textbf{0.8096183} & 0.8098543 & 0.8098237 & 0.8097445 \\
    $\gamma = 0.5$ $\delta = 1.5$ & 0.9287466 & \textbf{0.9284917} & 0.9288502 & 0.9287975 & 0.9286744 \\
    $\gamma = 1$ $\delta = 0.5$ & 0.5713549 & \textbf{0.5709173} & 0.5715235 & 0.5714033 & 0.5711274 \\
    $\gamma = 1$ $\delta = 1$ & 0.8084149 & \textbf{0.8075349} & 0.8088334 & 0.8085280 & 0.8079301 \\
    $\gamma = 1$ $\delta = 1.5$ & 0.9267936 & \textbf{0.9256779} & 0.9272953 & 0.9268617 & 0.9260850 \\
    \bottomrule
    \end{tabular}

\end{table}

\begin{table}[htb!]
    \centering
    \caption{MSE of different datasets from fitted tvSTAR($1_1$) model with Mexican hat wavelet and $J=3$.}
    \begin{tabular}{cccccc}
    \toprule
    %\multicolumn{6}{c}{Haar}\\ 
    %\bottomrule
    \multirow{2}*{Spatial weight matrix} & \multicolumn{2}{c}{$w^{di}_{ij}$} & \multicolumn{3}{c}{$w^{ne}_{ij}$}\\ \cline{2-6}
     &  $\alpha=0.5$ & $\alpha=1$ & $\alpha=0.5$ & $\alpha=1$ & $\alpha=2$ \\\hline
    $\gamma = 0.25$ $\delta = 0.5$ & 0.5713301 & 0.5713054 & 0.5713330 & 0.5713241 & \textbf{0.5713024} \\   
    $\gamma = 0.25$ $\delta = 1$ & 0.8092726 & \textbf{0.8092524}  & 0.8092760 & 0.8092697 & 0.8092562 \\ 
    $\gamma = 0.25$ $\delta = 1.5$ & 0.9283331 & \textbf{0.9283079}  & 0.9283398 & 0.9283333 & 0.9283203 \\
    $\gamma = 0.5$ $\delta = 0.5$ & 0.5712721 & \textbf{0.5711897} & 0.5712766 & 0.5712649 & 0.5712213 \\
    $\gamma = 0.5$ $\delta = 1$ & 0.8090920 & \textbf{0.8089110} & 0.8091462 & 0.8091135 & 0.8090265 \\
    $\gamma = 0.5$ $\delta = 1.5$ & 0.9280607 & \textbf{0.9278012} & 0.9281587 & 0.9281040 & 0.9279749 \\
    $\gamma = 1$ $\delta = 0.5$ & 0.5706313 & \textbf{0.5702010} & 0.5707714 & 0.5706620 & 0.5703890 \\
    $\gamma = 1$ $\delta = 1$ & 0.8076571 & \textbf{0.8067883} & 0.8080496 & 0.8077554 & 0.8071624 \\
    $\gamma = 1$ $\delta = 1.5$ & 0.9259958 & \textbf{0.9249419} & 0.9265289 & 0.9261045 & 0.9253332 \\
    \bottomrule
    \end{tabular}

\end{table}

\begin{table}[htb!]
    \centering
    \caption{MSE of different datasets from fitted tvSTARMA($1_1, 1_1$) model with Mexican hat wavelet and $J=2$.}
    \begin{tabular}{cccccc}
    \toprule
    %\multicolumn{6}{c}{Haar}\\ 
    %\bottomrule
    \multirow{2}*{Spatial weight matrix} & \multicolumn{2}{c}{$w^{di}_{ij}$} & \multicolumn{3}{c}{$w^{ne}_{ij}$}\\ \cline{2-6}
     &  $\alpha=0.5$ & $\alpha=1$ & $\alpha=0.5$ & $\alpha=1$ & $\alpha=2$ \\\hline
    $\gamma = 0.25$ $\delta = 0.5$ & 0.5493941 & \textbf{0.5493536} & 0.5494022 & 0.5493865 & 0.5493567 \\   
    $\gamma = 0.25$ $\delta = 1$ & 0.7960634 & \textbf{0.7960057} & 0.7960567 & 0.7960479 & 0.7960121 \\ 
    $\gamma = 0.25$ $\delta = 1.5$ & 0.9221163 & \textbf{0.9220534} & 0.9221938 & 0.9223765 & 0.9221010 \\
    $\gamma = 0.5$ $\delta = 0.5$ & 0.5493746 & \textbf{0.5492904} & 0.5493905 & 0.5493671 & 0.5493187 \\
    $\gamma = 0.5$ $\delta = 1$ & 0.7960146 & \textbf{0.7957764} & 0.7963313 & 0.7960042 & 0.7958850 \\
    $\gamma = 0.5$ $\delta = 1.5$ & 0.9217563 & \textbf{0.9214193} & 0.9218807 & 0.9217912 & 0.9215954 \\
    $\gamma = 1$ $\delta = 0.5$ & 0.5490224 & \textbf{0.5487275} & 0.5491266 & 0.5490394 & 0.5488448 \\
    $\gamma = 1$ $\delta = 1$ & 0.7948874 & \textbf{0.7941313} & 0.7952156 & 0.7949564 & 0.7944082 \\
    $\gamma = 1$ $\delta = 1.5$ & 0.9200001 & \textbf{0.9189014} & 0.9204669 & 0.9200260 & 0.9192314 \\
    \bottomrule
    \end{tabular}

\end{table}

\section{Application}

Precipitation is an important variable for climate and hydro-meteorology research and agricultural production. There are many studies of precipitation in different regions using time series models, such as Dalezios and Adamowski (1995), Wang et al. (2013) and Wu et al. (2021). In this paper, we use historical daily precipitation records obtained from GHCN (Global Historical Climatology Network)-Daily, an integrated public database of NOAA (National Oceanic and Atmospheric Administration), using R package rnoaa.
The data selected are daily precipitation (in tenths of millimeters) from Midwestern states of the USA. 
The region consists of 11 states: North Dakota, South Dakota, Illinois, Iowa, Kansas, Michigan, Minnesota, Missouri, Nebraska, Ohio and Wisconsin.
We use only climate monitoring stations which contains no missing data during the period between 1990-01-01 and 2009-12-30 (inclusive).
The 30 stations selected are presented in Figure 6 and Figure 7 shows precipitations recorded of the two weather stations.

\begin{figure}[h!]
    \centering
    \includegraphics[width=11.5cm]{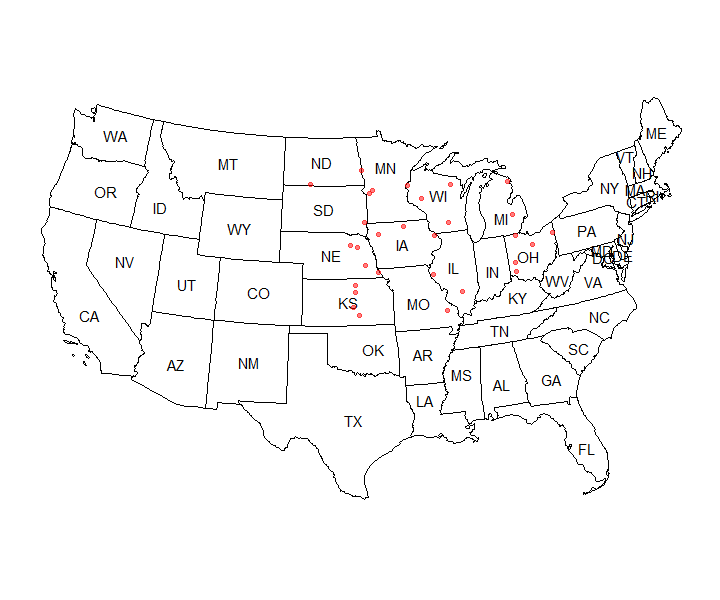} 
    \caption{Locations of the stations selected from Midwestern states of the USA.}
\end{figure}

\begin{figure}[ht!]
    \centering
    \includegraphics[width=12.5cm]{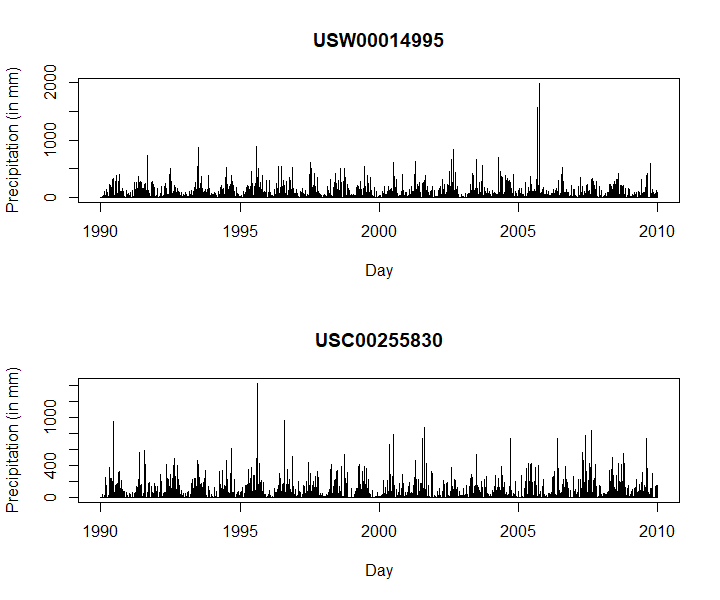}
    \caption{Precipitation recorded of 2 weather stations.}
    \label{fig:my_label}
\end{figure}

As precipitation data may have heavy tail distribution due to very large values with finite probabilities, we apply the transformation
\begin{equation}
    Z_i\left( \frac{t}{T} \right) = \log_{10}\left(Y_i\left( \frac{t}{T} \right)+1\right),
\end{equation}
where $Y_i\left( \frac{t}{T} \right)$ is the original series at location $\textbf{x}_i$ and time $t$ before applying the estimation proposed to remove the effects.

The inverse weight and negative exponential weight distances are selected in this application. The weights are defined as (61) and (62). We set $\alpha = 0.5, 1$ and 0.5, 1, 2 for inverse weight and negative exponential weight, respectively.

The results were obtained from the estimation procedure described in Sections 3.2 and 4.2 using tvSTAR($1_1$) and tvSTARMA($1_1,1_1$) models. The wavelet expansion of time-varying parameters is built using the Mexican hat wavelet with $J$ = 2, 3, 4 (the last value for tvSTAR models only).

Table 4 shows the MSEs of the estimates for different models with different spatial weight matrices selected. We can see that MSEs are better when the spatial weight matrix uses the inverse weights distance with $\alpha = 1$. Note that under the same fitted model, the MSEs are similar (same conclusion as in the simulations), therefore $J=2$ may be a satisfactory choice to fit the data, since it has less coefficients.

\begin{table}[htb]
    \centering
    \caption{MSE of different models where Mexican hat wavelet were used.}
    \begin{tabular}{cccccc}
    \toprule
    %\multicolumn{6}{c}{Mexican hat}\\ 
    %\bottomrule
    \multirow{2}*{Spatial weight matrix} & \multicolumn{2}{c}{$w^{di}_{ij}$} & \multicolumn{3}{c}{$w^{ne}_{ij}$} \\ \cline{2-6}
     & $\alpha=0.5$ & $\alpha=1$ & $\alpha=0.5$ & $\alpha=1$ & $\alpha=2$ \\\hline
    tvSTAR($1_1$) $J = 2$ & 1.142290e-4 & \textbf{1.140377e-4} & 1.148335e-4 & 1.143726e-4 & 1.141951e-4 \\   
    tvSTAR($1_1$) $J = 3$ & 1.142095e-4 & \textbf{1.140122e-4}  & 1.148155e-4 & 1.143552e-4 & 1.141754e-4\\ 
    tvSTAR($1_1$) $J = 4$ & 1.141900e-4 & \textbf{1.139864e-4}  & 1.147969e-4 & 1.143357e-4 & 1.141545e-4\\
    tvSTARMA($1_1,1_1$) $J = 2$ & 1.141545e-4 & \textbf{1.139164e-4} & 1.147790e-4 & 1.142976e-4 & 1.140730e-4\\
    tvSTARMA($1_1,1_1$) $J = 3$ & 1.141271e-4 & \textbf{1.138722e-4} & 1.147541e-4 & 1.142734e-4 & 1.140426e-4\\
    %tvSTARMA($1_1,1_1$) $J = 4$ & 1.141097e-4 & 1.138560e-4 & 1.147308e-4 & 1.147308e-4 & 1.140330e-4\\
    \bottomrule
    \end{tabular}

\end{table}

\section{Further comments}

Stationarity is a basic assumption in time series analysis, but sometimes it is difficult to guarantee it in practice. In this paper we proposed time-varying STARMA models based on the concept of locally stationary process in the sense of Dahlhaus. The time-varying parameters are expanded in terms of wavelet bases. Some simulations were performed and an application was given using the proposed estimation procedure. The Mexican hat wavelet was our choice, but some other bases could be used, for example Morlet, Shannon or Meyer.
For the time-varying STARMA models, there are many interesting studies that can be explored in a future research, for examples: (i) incorporate seasonal factors; (ii) explore innovation covariance function heterogeneity; (iii) consider the case when the spatial weight matrix is time-varying; (iv) compare the proposed model and other spatio-temporal models, such as spatial dynamic models and hierarchical spatio-temporal models; (v) investigate asymptotic properties of the tvSTARMA model, as was done for the tvSTAR
model in propositions 1 and 2.

\section*{Acknowledgements}

Yangyang Chen acknowledges the support of FAPESP, through grant 2019/05917-6. Pedro Alberto Morettin and Chang Chiann acknowledge the partial support of FAPESP grant 2018/04654-9.

\appendix
\numberwithin{equation}{section}
\section{Appendix}
\subsection{Proof of proposition 1}

Equation (23) can be written as
\begin{equation}
    \mat{\nu} \left( \frac{t}{T} \right) = \sum_{s=1}^p \sum_{l=0}^{\lambda_s} R_{sl} \left( \frac{t}{T} \right) W^{(l)} \textbf{Z} \left( \frac{t-s}{T} \right) + \mat{\epsilon}\left( \frac{t}{T} \right),
\end{equation}
where
\begin{equation}
    R_{sl} \left( \frac{t}{T} \right) = \sum_{j \ge J}^{\infty} \sum_{m = 0}^{2^j-1} \beta^{sl}_{j,m} \psi_{j,m}\left( \frac{t}{T} \right).
\end{equation}

Define 
\begin{equation}    
    S = \left[ \sum_{s=1}^p \sum_{l=0}^{\lambda_s} R_{sl} \left( \frac{p+1}{T} \right) W^{(l)} \textbf{Z} \left( \frac{p+1-s}{T} \right),...,\sum_{s=1}^p \sum_{l=0}^{\lambda_s} R_{sl} \left( \frac{T-p}{T} \right) W^{(l)} \textbf{Z} \left( \frac{T-p-s}{T} \right) \right]'
\end{equation}  
and
\begin{equation}    
    \mat{\epsilon} = \left[ \hbox{vec}\left(\mat{\epsilon}\left( \frac{p+1}{T} \right)\right)',...,\hbox{vec}\left(\mat{\epsilon}\left( \frac{T-p}{T} \right)\right) '\right]',
\end{equation}
equation (29) can be decomposed as
\begin{eqnarray}
    \hat{\mat{\beta}} &=& (\mat{\Psi}'\mat{\Psi})^{-1} \mat{\Psi}' (\mat{\Psi} \mat{\beta} + S + \mat{\epsilon}) \nonumber \\
    &=& \mat{\beta} + (E\mat{\Psi}'\mat{\Psi})^{-1} \mat{\Psi}' \mat{\epsilon} + [(\mat{\Psi}'\mat{\Psi})^{-1} - (E\mat{\Psi}'\mat{\Psi})^{-1}]\mat{\Psi}' \mat{\epsilon}  + (\mat{\Psi}'\mat{\Psi})^{-1}\mat{\Psi}' S \nonumber \\
    &=& \mat{\beta} + T_1 + T_2 + T_3.
\end{eqnarray}

Since $T_1 = (E\mat{\Psi}'\mat{\Psi})^{-1} \mat{\Psi}' \mat{\epsilon}$, we have $E[T_1] = 0$ and by Taylor expansion of the matrix $(\mat{\Psi}'\mat{\Psi})^{-1}$, $T_2 = T_{21} + T_{22}$, where
\begin{equation}
    T_{21} =  (E\mat{\Psi}'\mat{\Psi})^{-1}(E\mat{\Psi}'\mat{\Psi} - \mat{\Psi}'\mat{\Psi})(E\mat{\Psi}'\mat{\Psi})^{-1} \mat{\Psi}'\mat{\epsilon}
\end{equation}
and
\begin{equation}
    T_{22} = (E\mat{\Psi}'\mat{\Psi})^{-1}(E\mat{\Psi}'\mat{\Psi} - \mat{\Psi}'\mat{\Psi})(E\mat{\Psi}'\mat{\Psi})^{-1}(E\mat{\Psi}'\mat{\Psi} - \mat{\Psi}'\mat{\Psi})(E\mat{\Psi}'\mat{\Psi})^{-1} \mat{\Psi}'\mat{\epsilon}.
\end{equation}

Analogously to Chang and Morettin (2005), we obtain
\begin{eqnarray}
    \norm{T_{21}}_2 &\le& \norm{(E\mat{\Psi}'\mat{\Psi})^{-1}}^2_2 \norm{\hbox{Cov}(\mat{\Psi}'\mat{\Psi})}^{1/2}_2 \norm{\hbox{Cov}(\mat{\Psi}'\mat{\epsilon})}^{1/2}_2 \nonumber \\
    &=& O((nT)^{-2}) O(2^{2J}(nT)^{1/2}) O((nT)^{1/2}) \nonumber \\
    &=& O(2^J (nT)^{-1})
\end{eqnarray}
and
\begin{eqnarray}
    \norm{T_{22}}_2 &\le& \norm{(E\mat{\Psi}'\mat{\Psi})^{-1}}^3_2 \norm{\hbox{Cov}(\mat{\Psi}'\mat{\Psi})}^2_2 \norm{\hbox{Cov}(\mat{\Psi}'\mat{\epsilon})}_2 \nonumber \\
    &=& O((nT)^{-3}) O(2^{2J}nT) O((nT)^{1/2}) \nonumber \\
    &=& O(2^{2J}(nT)^{-3/2}).
\end{eqnarray}

From Donoho et al. (1995), we have
\begin{equation}
    \hbox{sup} \left\{\sum_{j \ge J}^{\infty} \sum_{k = 0}^{2^j-1} |\beta^{sl}_{j,k}|^2 \right\} = O(2^{-2Ju}),
\end{equation}
where $u=d+1/2-1/\Tilde{p}$ with $\Tilde{p}=\hbox{min}\{p,2\}$ and similarly to Dahlhaus et al. (1999),
\begin{eqnarray}
    \norm{T_3}_2 &\le& (\norm{(E\mat{\Psi}'\mat{\Psi})^{-1}}_2 + \norm{(\mat{\Psi}'\mat{\Psi})^{-1} - (E\mat{\Psi}'\mat{\Psi})^{-1}}_2 ) \norm{\mat{\Psi}' S}_2 \nonumber \\
    &=& (O((nT)^{-1}) + O(2^J(nT)^{-3/2}) + O(2^{2J}(nT)^{-2})) \norm{\mat{\Psi}' S}_2 \nonumber \\
    &=& O((nT)^{-1})O(nT(2^{-Ju} + (nT)^{-1/2} 2^{-J(d-1/2-1/(2\Tilde{p}))})\sqrt{\log{(nT)}})) \nonumber \\
    &=& O((2^{-Ju} + (nT)^{-1/2} 2^{-J(d-1/2-1/(2\Tilde{p}))})\sqrt{\log{(nT)}})) \nonumber \\
    &=& O((nT)^{-1/2-\tau}),
\end{eqnarray}
for some $\tau > 0$. Then, $\norm{T_3}_2 \le O((nT)^{-1/2})$.

The result $(i.)$ follows.

For $(ii.)$, we have
\begin{eqnarray}
    E[T_1T_1'] &=&(E\mat{\Psi}'\mat{\Psi})^{-1} \hbox{Cov}(\mat{\Psi}'\mat{\epsilon})(E\mat{\Psi}'\mat{\Psi})^{-1} \nonumber \\
    & \le & \norm{(E\mat{\Psi}'\mat{\Psi})^{-1}}_2^2 \norm{\hbox{Cov}(\mat{\Psi}'\mat{\epsilon})}_2 \nonumber \\
    &=& O((nT)^2)O(nT) \nonumber \\
    &=& O((nT)^{-1}).
\end{eqnarray}

Analogously,
\begin{eqnarray}
    E[T_2T_2'] &=& O(2^{2J}(nT)^{-2}) + O(2^{4J}(nT)^{-3}), \\
    E[T_3T_3'] &=& O((nT)^{-1}), \\
    E[T_1T_2'] &=& E[T_1'T_2] = E[T_2T_3'] = E[T_2'T_3] = O(2^J (nT)^{-3/2}) + O(2^{2J} (nT)^{-2}),\\
    E[T_1T_3'] &=& E[T_1'T_3] = O((nT)^{-1}),
\end{eqnarray}
and then the result $(ii.)$ follows.

\subsection{Proof of proposition 2}

\begin{lemma}
    \normalfont
    Suppose the assumptions A.1-A.4 hold, then
    \begin{equation}
        N\textbf{H}(\mat{\Psi}'\mat{\Psi})^{-1}\textbf{H}' \mathop\to^P \mat{\Gamma},
    \end{equation}
    where $\mathbf{\Psi}$ is as in (27) and
    \begin{equation}
        \mat{\Gamma} = \lim_{N \to \infty} N\textbf{H}E[\mat{\Psi}'\mat{\Psi}]^{-1}\textbf{H}'.
    \end{equation}
    
\end{lemma}

\textbf{Proof.} Let $\psi_{gl}$ the $g$th row and $l$th column cell of $\mat{\Psi}$, we have
\begin{equation}
    \psi_{gl} = \sum_{i=1}^n f^i_{gl} Z^i_{gl},
\end{equation}
where $f^i_{gl}$ is obtained by the product of a wavelet function and a weight function up to one and $Z^i_{gl}$ is a locally stationary process, both are elements of $(g,l)$th cell of $\mat{\Psi}$ at location $i$.

Let $(\psi\psi)_{lm}$ the $l$th row and $m$th column cell of any subpartition of $\mat{\Psi}'\mat{\Psi}$, $(\psi\psi)_{lm}$ can be expressed as
\begin{eqnarray*}
    (\psi\psi)_{lm} &=& \sum_{g=1}^N \psi_{gl} \psi_{gm} \\
    &=& \sum_{g=1}^N \left[\sum_{i=1}^n(f^i_{gl} Z^i_{gl}) \sum_{i=1}^n(f^i_{gm} Z^i_{gm})\right] \\
    &=& \sum_{g=1}^N f^1_{gl} Z^1_{gl} f^1_{gm} Z^1_{gm} + \sum_{g=1}^N f^1_{gl} Z^1_{gl} f^2_{gm} Z^2_{gm} + \dots + \sum_{g=1}^N f^n_{gl} Z^n_{gl} f^n_{gm} Z^n_{gm}.
\end{eqnarray*}

The proof of $(\psi'\psi)_{lm}$ is similar to that of Sato et al. (2007) and the result follows from the Weak Law for $L^1-Maxingales$ of Andrews (1988).

\begin{lemma}
    \normalfont
    Suppose the assumptions A.1-A.4 hold, then
    \begin{equation}
        \frac{1}{\sqrt{N}} \textbf{G} (\mat{\Psi}'\mat{\epsilon}) \mathop\to^D \mathscr{N}_K(\textbf{0}, \mat{\Gamma}^{-1}),
    \end{equation}
    where $\mat{\epsilon}$ is as in (A.4) and covariance matrix 
    \begin{equation}
        \mat{\Gamma}^{-1} = \lim_{N \to \infty} \textbf{G} \frac{E[\mat{\Psi}'\mat{\Psi}]}{N} \textbf{G}',
    \end{equation}
    where $N = n(T-p)$, \textbf{G} is a matrix with the same number of columns of $\mat{\Psi}$ and $K$ rows.
\end{lemma}

\textbf{Proof.} Let $(\psi e)_l$ the $l$th element of $\mat{\Psi}'\mat{\epsilon}$, 
\begin{eqnarray*}   
    (\psi e)_l &=& \sum_{g=1}^N (\psi e)_{gl} \\
    &=& \sum_{g=1}^N f_{gl} \Tilde{Z}_{gl} \epsilon_g,
\end{eqnarray*}
where $f_{gl}$ is a wavelet function and $\Tilde{Z}_{gl}$ is a weighted arithmetic mean of $n$ locally stationary processes with weights sum to one, both are of $(g,l)$th cell of $\mat{\Psi}$ and $\epsilon_g$, $g$th element of $\mat{\epsilon}$, is independent Gaussian process.

Let $\mat{\Psi}'\mat{\epsilon} = \sum_{g=1}^N b_g$, where $b_g$ is a vector consist of $(\psi e)_{gl}$ and define $\mathscr{F}_g$ as a $\sigma$-algebra containing all the information up to moment $g$. Since
\begin{equation}
    E[b_g \mid \mathscr{F}_{g-1}] =0,
\end{equation}
$\{ b_g\}_{g=1}^{\infty}$ is a martingale difference sequence. Then
\begin{eqnarray*}
    E \left[ \frac{\sum_{g=1}^N b_g b_g'}{N} \right] &=& E \left[ \frac{\mat{\Psi}'\mat{\epsilon} \mat{\epsilon}' \mat{\Psi}}{N} \right] \\
    &=& E \left[ E\left[ \frac{\mat{\Psi}'\mat{\epsilon} \mat{\epsilon}' \mat{\Psi}}{N} \middle\vert\ \mat{\Psi} \right] \right] \\
    &=& E\left[ \frac{\mat{\Psi}'\mat{\Psi}}{N} \right].
\end{eqnarray*}

Analogously, 
\begin{equation}
    \textbf{G} E \left[ \frac{\sum_{g=1}^N b_g b_g'}{N} \textbf{G}' = \textbf{G} E\left[ \frac{\mat{\Psi}'\mat{\Psi}}{N} \right] \right] \textbf{G}' \to \mat{\Gamma}^{-1},
\end{equation}
where
\begin{equation}
    \mat{\Gamma}^{-1} = \lim_{N \to \infty} \textbf{G} \frac{E[\mat{\Psi}'\mat{\Psi}]}{N} \textbf{G}'
\end{equation}
and similar to Lemma 1, 
\begin{equation}
    \textbf{G}\frac{\sum_{g=1}^N b_g b_g'}{N} \textbf{G}' \mathop\to^P \mat{\Gamma}^{-1}.
\end{equation}

The result follows the central limit theorem for martingale difference sequence of White (2000).

From equation (A.11), $\sqrt{N}\norm{T_3}_2 = O((nT)^{-\tau}) = o_p(1)$, applying Lemma 1, Lemma 2 and Slutsky theorem, we have
\begin{eqnarray}
    \sqrt{N} \textbf{H} (\hat{\mat{\beta}} -\mat{\beta}) &=& \sqrt{N} \textbf{H} (T_1+T_2+T_3) \\
    &=& \sqrt{N} \textbf{H} (T_1+T_2) + o_p(1) \\
    &=& \sqrt{N} \textbf{H}(\mat{\Psi}'\mat{\Psi})^{-1}\mat{\Psi}'\mat{\epsilon} + o_p(1) \\
    &=& \mathscr{N} +o_p(1),
\end{eqnarray}
where $\mathscr{N}$ has a $K$-dimensional normal distribution with mean zero and covariance matrix $\mat{\Gamma}$ as in (A.18). The result follows the Slutsky's theorem.

\section*{References}
\addcontentsline{toc}{chapter}{\protect\numberline{}References}%

\begin{enumerate}

\item Andrews, D. W. K. (1988) Laws of large number for dependent non-identically distributed random variables, \textit{Econometric Theory}, \textbf{4}(3), 458-467.

\item Box, G. E. P. and Jenkins, G. M. (1970) \textit{Time series Analysis: Forecasting and Control}, San Francisco: Holden-Day.

\item Chiann, C. and Morettin, P. A. (1999) Estimation of time-varying linear systems, \textit{Statistical Inference for Stochastic Processes} \textbf{2}, 253–285. %https://doi.org/10.1023/A:1009999208631

\item Chiann, C. and Morettin, P. A. (2005) Time domain nonlinear estimation of time varying linear systems, \textit{Journal of Nonparametric Statistics}, \textbf{17}(3), 365–383.

\item Cipra, T. and Motykova, I. (1987) Study on Kalman filter in time series analysis, \textit{Commentationes Mathematicae Universitatis Carolinae}, \textbf{28}(3):549-563.

\item Cliff, A. D. and Ord, J. K. (1975) Space-Time Modeling with an Application to Regional Forecasting, \textit{Transactions of the Institute of British Geographers}, No.64, 119-128.

\item Dahlhaus, R. (1996a) Maximum likelihood estimation and model selection for locally stationary processes, \textit{Journal of Nonparametric Statistics}, \textbf{6}(2-3), 171-191.
%doi: 10.1080/10485259608832670. URL http://dx.doi.org/10.1080/10485259608832670. 

\item Dahlhaus, R. (1996b) On the kullback-leibler information divergence of locally stationary processes, \textit{Stochastic Processes and their Applications}, \textbf{62}(1), 139–168. 
%ISSN 0304-4149. %doi: https://doi.org/10.1016/0304-4149(95)00090-9. URL http://www.sciencedirect.com/science/article/pii/0304414995000909. 

\item Dahlhaus, R. (1996c) Asymptotic statistical inference for nonstationary processes with evolutionary spectra, In: Robinson, P. M., Rosenblatt, M. (eds) \textit{Athens Conference on Applied Probability and Time Series Analysis. Volume II: Time Series Analysis In Memory of E.J. Hannan}, \textbf{115}, 145–159. Springer, New York, NY.
%ISBN 978-1-4612-2412-9. doi: 10.1007/978-1-4612-2412-9 11. URL https://doi.org/10.1007/978-1-4612-2412-9{ }11. 

\item Dahlhaus, R. (1997) Fitting time series models to nonstationary processes, \textit{The Annals of Statistics}, \textbf{25}(1), 1–37.

\item Dahlhaus, R. (2000) A likelihood approximation for locally stationary processes, \textit{The Annals of Statistics}, \textbf{28}(6), 1762–1794.

\item Dahlahus, R. (2012) Locally stationary processes, \textit{Time Series Analysis: Methods and Applications}, \textbf{30}, 351-413.

\item Dahlhaus, R., Neumann, M. H. and von Sachs, R. (1999) Nonlinear wavelet estimation of time-varying autoregressive processes, \textit{Bernoulli} \textbf{5}(5), 873–906.

\item Dalezios, N. R and Adamowski, K. (1995) Spatio-temporal precipitation modelling in rural watersheds, \textit{Hydrological Sciences Journal}, \textbf{40}(5), 553-568. 

\item Donoho, D., Johnstone, I., Kerkyacharian, G., Picard, D. (1995) Wavelet shrinkage: asymptopia? \textit{Journal of the Royal Statistical Society}, Series B (Methodological), \textbf{57}(2), 301-337.

\item Hu, X., Qin, Z. and Chu, F. (2011) Damage detection in plate
structures based on spacetime autoregressive moving average
processes, \textit{Journal of Physics: Conference Series}, \textbf{305}(1):012119.

\item Jin, E. Y. (2017) Estrutura de vizinhanças espaciais nos modelos autorregressivos e de médias móveis espaço-temporais STARMA, dissertação de mestrado, Universidade de São Paulo Instituto de Matemática e Estatística, São Paulo.

\item Kalman, R. (1960) A New Approach to Linear Filtering and Prediction Problems, \textit{Journal of Basic Engineering}, \textbf{82}(1), 35-45. 

\item Kamarianakis, Y. and Prastacos, P. (2005) Space–time modelling of traffic flow,  \textit{Computers \& Geosciences}, \textbf{31}(2), 119-133.

\item Kurt, S. and Tunay, K. B. (2015) STARMA Models Estimation with Kalman Filter: The Case of Regional Bank Deposits. \textit{Procedia - Social and Behavioral Sciences}, Vol. 195, 2537-2547.  %http//doi.org/10.1016/j.sbspro.2015.06.441 

\item Martin, R. L. and Oeppen, J. E. (1975) The identification of regional forecasting models using space: time correlation functions,  \textit{Transactions of the Institute of British Geographers}, No. 66, 95-118.

\item Pace, R. K., Barry, R., Gilley, O. W. and Sirmans, C. F. (2000) A method for spatial-temporal forecasting with an application to real estate prices, \textit{International Journal of Forecasting}, \textbf{16}(2), 229-246.

\item Pfeifer, P. E., and Deutsch, S. J. (1980a) A three-stage iterative procedure for space-time modeling, \textit{Technometrics}, \textbf{22}(1), 35-47. 

\item Pfeifer, P. E., and Deutsch, S. J. (1980b) Identification and Interpretation of First-Order Space-Time ARMA Models, \textit{Technometrics}, \textbf{22}(3), 397-408.

\item Pfeifer, P. E., and Deutsch, S. J. (1981a) Variance of the Sample-Time Autocorrelation Function of Contemporaneously Correlated Variables, \textit{SIAM Journal of Applied Mathematics}, \textbf{40}(1), 133-136. 

\item Pfeifer, P. E., and Deutsch, S. J. (1981b) Seasonal Space-Time ARIMA modeling, \textit{Geographical Analysis}, \textbf{13}(2), 117-133. 

\item Pfeifer, P. E., and Deutsch, S. J. (1981c) Space-Time ARMA Modeling with contemporaneously correlated innovations, \textit{Technometrics}, \textbf{23}(4), 410-409.

\item Rohan, N. and Ramanathan T. V. (2013) Nonparametric estimation of a time-varying GARCH model, \textit{Journal of Nonparametric Statistics}, \textbf{25}(1), 33-52.

\item Sato, J. R., Morettin, P. A., Arantes, P. R. and Amaro Jr, E. (2007) Wavelet based time-varying vector autoregressive modelling, \textit{Computational Statistics \& Data Analysis}, \textbf{51}(12),  5847-5866.

\item Wang, S., Feng, J. and Liu, G. (2013) Application of seasonal time series model in the precipitation forecast, \textit{Mathematical and Computer Modelling}, \textbf{58}(3–4), 677-683.

\item White, H., (2000) Asymptotic Theory for Econometricians, revised edition, Academic Press, New York.

\item Wu, X., Zhou, J., Yu, H., Liu, D., Xie, K., Chen, Y., Hu, J., Sun, H. and Xing, F. (2021) The Development of a Hybrid Wavelet-ARIMA-LSTM Model for Precipitation Amounts and Drought Analysis, \textit{Atmosphere}, \textbf{12}(1):74. 

\item Yousuf, K. and Ng, S. (2021) Boosting high dimensional predictive regressions with time varying parameters, \textit{Journal of Econometrics}, \textbf{224}(1), 60-87. 

\item Zou, J., Zhu, J., Xie, P., Xuan, P. and Lai, X. (2018) A STARMA Model for Wind Power Space-Time Series. \textit{ 2018 IEEE Power \& Energy Society General Meeting (PESGM)}, 1-5.

\end{enumerate}

\end{document}